\documentclass[sigconf]{acmart}

\copyrightyear{2026}
\acmYear{2026}
\setcopyright{cc}
\setcctype{by}
\acmConference[WWW '26] {Proceedings of the ACM Web Conference 2026}{April 13--17, 2026}{Dubai, United Arab Emirates.}
\acmBooktitle{Proceedings of the ACM Web Conference 2026 (WWW '26), April 13--17, 2026, Dubai, United Arab Emirates}
\acmISBN{979-8-4007-2307-0/2026/04}
\acmDOI{10.1145/XXXXXX.XXXXXX}
\usepackage[dvipsnames]{xcolor}
\usepackage{booktabs}
\usepackage{mathrsfs}
\usepackage{xspace}
\usepackage{amsthm}
\usepackage{tcolorbox}
\usepackage{algorithm}
\usepackage{algpseudocode}
\usepackage{array}
\usepackage{multirow}
\usepackage{color, colortbl}
\definecolor{Gray}{gray}{0.9}
\usepackage{fontawesome5}
\usepackage[compatibility=false]{caption}
\usepackage[font=footnotesize,labelfont=rm,textfont=rm]{subcaption}
\usepackage{makecell}
\usepackage{textcomp}
\usepackage{url}
\usepackage{verbatim}
\usepackage{graphicx}
\usepackage{pifont}

\newcommand{\ourmethod}{\textsc{OFA-MAS}\xspace}
\usepackage{enumitem}

\begin{document}

\title{OFA-MAS: One-for-All Multi-Agent System Topology Design based on Mixture-of-Experts Graph Generative Models}

\author{Shiyuan Li}
\orcid{0000-0002-4381-7497}
\authornote{Both authors contributed equally to the paper}
\affiliation{%
  \institution{Griffith University}
  \city{Gold Coast}
  \country{Australia}}
  \email{shiyuan.li@griffithuni.edu.au}

\author{Yixin Liu}
\orcid{0000-0002-4309-5076}
\authornotemark[1]
\affiliation{%
  \institution{Griffith University}
  \city{Brisbane}
  \country{Australia}}
  \email{yixin.liu@griffith.edu.au}

\author{Yu Zheng}
\orcid{0000-0003-0757-4210}
\affiliation{%
  \institution{Griffith University}
  \city{Gold Coast}
  \country{Australia}}
  \email{yu.zheng@griffith.edu.au}

\author{Mei Li}
\orcid{0000-0003-2962-8945}
\affiliation{%
  \institution{Northwest A\&F University}
  \city{Shaanxi}
  \country{China}}
  \email{limei@nwsuaf.edu.cn}

\author{Quoc Viet Hung Nguyen}
\orcid{0000-0002-9687-1315}
\affiliation{%
  \institution{Griffith University}
  \city{Gold Coast}
  \country{Australia}}
  \email{henry.nguyen@griffith.edu.au}

\author{Shirui Pan}
\orcid{0000-0003-0794-527X}
\authornote{Corresponding Author}
\affiliation{%
  \institution{Griffith University}
  \city{Gold Coast}
  \country{Australia}}
  \email{s.pan@griffith.edu.au}

\renewcommand{\shortauthors}{Li et al.}

\begin{abstract}
Multi-Agent Systems (MAS) offer a powerful paradigm for solving complex problems, yet their performance is critically dependent on the design of their underlying collaboration topology. As MAS become increasingly deployed in web services (e.g., search engines), designing adaptive topologies for diverse cross-domain user queries becomes essential. Current graph learning-based design methodologies often adhere to a ``one-for-one'' paradigm, where a specialized model is trained for each specific task domain. This approach suffers from poor generalization to unseen domains and fails to leverage shared structural knowledge across different tasks. To address this, we propose \ourmethod, a one-for-all framework that generates adaptive collaboration graphs for any task described in natural language through a single universal model. Our approach integrates a Task-Aware Graph State Encoder (TAGSE) that filters task-relevant node information via sparse gating, and a Mixture-of-Experts (MoE) architecture that dynamically selects specialized sub-networks to drive node and edge prediction. We employ a three-stage training strategy: unconditional pre-training on canonical topologies for structural priors, large-scale conditional pre-training on LLM-generated datasets for task-topology mappings, and supervised fine-tuning on empirically validated graphs. Experiments across six diverse benchmarks show that \ourmethod significantly outperforms specialized one-for-one models, generating highly adaptive MAS topologies. Code: https://github.com/Shiy-Li/OFA-MAS.
\end{abstract}



\received{20 February 2007}
\received[revised]{12 March 2009}
\received[accepted]{5 June 2009}

\maketitle

\section{Introduction}
Multi-agent systems (MAS) built upon large language models have emerged as a powerful approach for solving complex problems by orchestrating multiple specialized agents with distinct roles~\cite{hong2024metagpt,pan2026correcting,pan2025explainable,miao2025blindguard}. To date, MAS have been increasingly integrated with web services, such as search engines, online knowledge bases, and cloud-based task execution platforms~\cite{zhuge2024gptswarm,yao2023react,shinn2023reflexion,li2026clip}. However, the effectiveness of MAS highly depends on their \textit{collaboration topology}, i.e., the structural design determining which agents communicate, when they interact, and how knowledge propagates~\cite{qian2024scaling-macnet,zhang2025g,wu2024autogen}. An optimal topology applies the right expertise at the right moment, while poor topology design can lead to coordination inefficiencies that undermine performance. 

To address this challenge of topology design, early efforts focus on manual designs like sequential chains~\cite{wei2022chain,zhang2022automatic}, tree hierarchies~\cite{yao2023tree}, and debate structures have shown domain-specific success~\cite{du2023improving,liu2023dylan}. 
However, such manually crafted structures are often rigid and hard to generalize, since a topology that works well in one domain may fail to adapt to new tasks~\cite{liu2025graph,li2026assemble}. 
More recently, \textit{automatic topology design} emerged as a promising solution, which aims to adaptively construct agent interaction structures for a given query using deep graph learning models. For example, a line of methods constructs topologies by pruning predefined dense graphs~\cite {zhang2024agentprune,wang2025agentdropout}. Another line of methods leverages GNN message passing~\cite{li2024noise,liu2024arc,pan2025survey} to adaptively learn task-specific interaction topologies for collaboration graphs~\cite{zhuge2024gptswarm,zhang2025g,shen2025understanding,chen2025uncertainty,zhao2025freegad}.

Despite the effectiveness of the existing graph-based automatic topology design methods, these methods mainly focus on a ``\textbf{one-for-one}'' learning paradigm, that is, a topology design model is trained on a specific task dataset (e.g., MMLU) and can only generalize within these tasks. Such a paradigm, unfortunately, introduces critical barriers to the practical and large-scale deployment of these methods, which can be characterized by three shortcomings: (1)~\textbf{Impractical Domain Assumptions}: The one-for-one paradigm presumes optimization for a single and pre-defined task domain, yet real-world agentic systems in web services face user queries spanning unpredictable domains with cross-domain reasoning. Since general-purpose MAS cannot reasonably expect users to pre-categorize their requests, the domain constraint limits the application of the current methods to real-world user-facing services. 
(2)~\textbf{Intractable Scalability and Maintenance}: While training multiple specialist models for each new domain may alleviate the domain limitation, its prohibitive overhead that scales with the increasing number of supported tasks compounds the difficulty. Each domain requires a dedicated pipeline for data collection, training, and tuning, rendering the approach unsustainable for broad applications, which creates an intractable maintenance burden in practical deployment. 
(3)~\textbf{Overlooking Cross-Domain Shared Knowledge}: Learning one-for-one design models overlooks the opportunity to discover and leverage shared design patterns across domains. For instance, domains like mathematical theorem proving and software debugging often share the same underlying collaboration patterns (e.g., a ``Analyst → Inspector → Solver'' workflow). Training isolated models may hinder us from learning such abstract and reusable problem-solving knowledge, leading to redundant training efforts and suboptimal performance. 

\begin{figure}[t]
    \centering
    \begin{subfigure}[b]{0.4\textwidth}
        \centering
        \includegraphics[width=\textwidth]{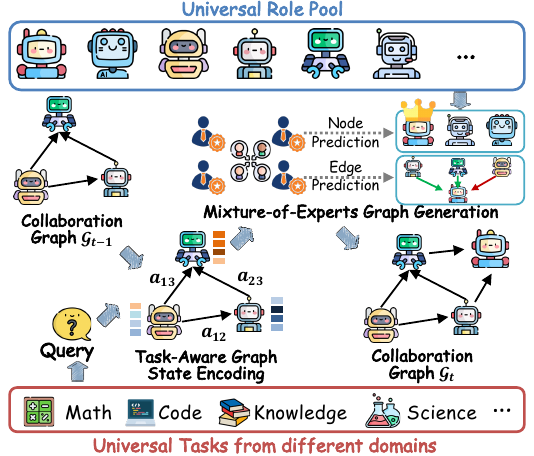}
        \caption{Architecture of the generative model of \ourmethod.}
        \label{fig:intro_1}
    \end{subfigure}
    \begin{subfigure}[b]{0.4\textwidth}
        \centering
        \includegraphics[width=\textwidth]{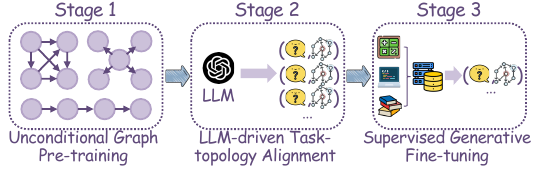}
        \caption{Pipeline of the training procedure of \ourmethod.}
        \label{fig:intro_2}
    \end{subfigure}   
    \caption{Skecth maps of \ourmethod.}
    \label{fig:intro}
    
\end{figure}

To overcome these limitations, a promising solution is to develop a \textbf{``one-for-all''} approach, designing a universal topology designer capable of generating optimal collaboration topologies across any task domain.  A one-for-all design approach can naturally break the domain-specific limitation due to its generalization capability. Moreover, it can also increase scalability in practical scenarios by only maintaining a single universal topology designer instead of training multiple domain-specific models. More importantly, the underlying shared knowledge across domains can be exploited to improve problem-solving efficiency.

In spite of its promising potential, developing a one-for-all method is a non-trivial task, facing two critical challenges. 
\textbf{Challenge 1 - Model Architecture Design}: Current topology design models typically rely on predefined and domain-specific role pools and lack mechanisms to incorporate task semantics into role selection. As a result, they often yield sub-optimal agent configurations when task demands differ across domains. Furthermore, traditional one-shot generation mechanisms hinder effective specialization, potentially leading to degraded performance when adapted to diverse task types. In this case, the architecture of one-for-all models should incorporate task-aware encoding mechanisms to capture global task context during topological design, while deploying an expert specialization framework that dynamically activates domain-specific expertise and preserves cross-task knowledge sharing. 
\textbf{Challenge~2 - Training Strategy}: To train a universal model for topology design, a prerequisite is large-scale and high-quality training data from multiple domains. However, collecting sufficient task queries with ground-truth optimal topologies is prohibitively expensive and labor-intensive. Moreover, even if such data were available, naively training on mixed-domain data could lead to catastrophic forgetting, where learning new domains degrades performance on previously learned ones. Therefore, an advanced training strategy should address the limitation of scarce data and design a learning curriculum that promotes effective cross-domain generalization.

Aiming to address the above challenges, we propose \ourmethod, a unified graph generative framework that combines task-adaptive model architecture with a multi-stage training strategy for universal MAS topology design. To overcome \textbf{Challenge 1}, we develop an autoregressive graph generation model for MAS design with two designs (See Figure~\ref{fig:intro_1}). To extract task-related information to guide topology design, we propose a \textit{task-aware graph state encoder} (TAGSE) that employs context-gated message passing to filter and regulate node representations by task relevance. Meanwhile, we design a \textit{mixture-of-experts (MoE) graph generation module} to capture diverse collaboration patterns across domains with multiple experts, routed by an expert gating network to contribute to topology generation. 
To address \textbf{Challenge 2}, we design an \textit{easy-to-hard three-stage training strategy} consisting of unconditional pre-training, LLM-guided conditional training, and supervised fine-tuning (See Figure~\ref{fig:intro_2}). The well-designed learning curriculum gradually introduces task consistency and preserves structural generation knowledge, mitigating catastrophic forgetting and enhancing cross-domain generalization. To mitigate data limitation, we propose a LLM-driven data synthesis pipeline that constructs multi-domain ``query–topology'' pair data with reduced annotation cost.
Overall, our work makes the following contributions:
\begin{itemize}[leftmargin=*]
    \item \textbf{Practical Paradigm}: We pioneer the transition from the ``one-for-one'' to ``one-for-all'' paradigm in graph-based MAS topology design, establishing the first universal framework for cross-domain collaboration graph generation.
    \item \textbf{Novel Method}: 
    We propose \ourmethod, an autoregressive graph generation framework with TAGSE and a MoE module, trained via a three-stage curriculum with LLM-driven data synthesis.
    \item \textbf{Sufficient Experiments}: 
    Experiments on six multi-domain benchmarks demonstrate that \ourmethod outperforms existing specialized methods with strong generalization, significantly advancing universal autonomous agent systems.
\end{itemize}


\section{Problem Formulation}
\subsection{Formulate MAS as Graphs}
A MAS constitutes a collaborative network of autonomous agents that collectively solve complex problems through structured interactions and coordinated reasoning. In the context of LLMs, we formalize a MAS as a structured collaboration network $\mathcal{G} = (\mathcal{V}, \mathcal{E})$, where the topology explicitly defines both the participating agents and their interaction patterns. Each vertex $v_i \in \mathcal{V}$ corresponds to an agent instantiated from an LLM, characterized by two key attributes: a \textit{functional role} $\rho_i$ that determines its specialized expertise (e.g., \texttt{Analyst}, \texttt{Programming Expert}, \texttt{Inspector}), and a \textit{contextual state} $\sigma_i$ that encapsulates its accumulated knowledge from previous interactions. The edge set $\mathcal{E} \subset \mathcal{V} \times \mathcal{V}$ encodes the information flow topology as a directed acyclic graph (DAG), where each edge $(v_j, v_i) \in \mathcal{E}$ establishes a communication channel from agent $v_j$ to agent $v_i$. We denote the set of immediate information providers for agent $v_i$ as $\Psi(v_i) = \{v_j \mid (v_j, v_i) \in \mathcal{E}\}$.

The execution protocol follows a topologically-ordered activation sequence over multiple communication rounds, ensuring information dependencies are respected. The system can execute for $K$ rounds to enable iterative refinement. In round $k$, agent $v_i$ generates its message $\omega_i^{(k)}$ by invoking its underlying language model:
\begin{equation}
\omega_i^{(k)} = \text{LLM}_i(\text{Prompt}(\rho_i, \sigma_i, \mathcal{Q}, \{\omega_j^{(k-1)} \mid v_j \in \Psi(v_i)\})).
\end{equation}
The system's final output aggregates the final-round messages according to a task-appropriate combination strategy.

\subsection{From One-for-One to One-for-All Paradigm}
\noindent\textbf{One-for-One MAS Design.} Traditional graph-based MAS topology design~\cite{zhang2025g,shen2025understanding} follows a domain-specific paradigm where individual design models are developed for single task domains. Formally, given a task domain $\mathcal{D}$ characterized by a distribution of queries $\mathcal{Q} \sim \mathcal{D}$ and a predefined role set $\mathcal{R}_{\mathcal{D}}$, the goal is to learn a domain-specific mapping function $f_{\mathcal{D}}: (\mathcal{Q}, \mathcal{R}_{\mathcal{D}}) \rightarrow \mathcal{G}$ that generates optimal collaboration topologies for queries within that domain. In the conventional setting of ``one model for one domain'', the design model $f_{\mathcal{D}}$ is optimized exclusively on the target domain dataset $\mathcal{T}_{\mathcal{D}} = \{(\mathcal{Q}_i, \mathcal{G}_i^*)\}_{i=1}^{|\mathcal{T}_{\mathcal{D}}|}$, where each pair $(\mathcal{Q}_i, \mathcal{G}_i^*)$ represents a query and its corresponding optimal topology. After sufficient training on $\mathcal{T}_{\mathcal{D}}$, the specialized model $f_{\mathcal{D}}$ can generate effective topologies for new queries from the same domain $\mathcal{D}$ during inference.
While effective within individual domains, this approach has inherent limitations in cross-domain transferability and scalability.

\noindent\textbf{One-for-All (OFA) MAS Design.} To address these limitations, we propose the \textbf{one-for-all graph-based MAS design}, where a single model capable of generating optimal collaboration topologies for diverse task domains without requiring domain-specific re-training. Formally, let $\mathcal{D}_{\text{train}} = \{\mathcal{D}_1, \mathcal{D}_2, \ldots, \mathcal{D}_N\}$ be a collection of training domains, where each domain $\mathcal{D}_i$ (with $i \in \{1, \ldots, N\}$) is associated with training and testing splits: $\mathcal{T}_i^{\text{train}} = \{(\mathcal{Q}_j^{(i)}, \mathcal{G}_j^{*(i)})\}_{j=1}^{|\mathcal{T}_i^{\text{train}}|}$ and $\mathcal{T}_i^{\text{test}}$, where $j$ indexes the query-topology pairs within each domain. We aim to train an OFA designer $f_{\text{OFA}}$ on the combined multi-domain training data $\mathcal{T}_{\text{train}} = \bigcup_{i=1}^{N} \mathcal{T}_i^{\text{train}}$, so that $f_{\text{OFA}}$ can generate effective topologies for queries across all test splits $\mathcal{T}_{\text{test}} = \bigcup_{i=1}^{N} \mathcal{T}_i^{\text{test}}$, where $\mathcal{T}_{\text{train}} \cap \mathcal{T}_{\text{test}} = \emptyset$.

Unlike traditional one-for-one graph-based approaches that require separate models for each domain, our OFA formulation enables a single model to handle diverse task types simultaneously. During inference, the model can generate appropriate topologies for any query from the supported domains without requiring domain-specific fine-tuning or model selection.

\section{Methodology}
\begin{figure}[t]
  \centering
    \includegraphics[width=1\linewidth]{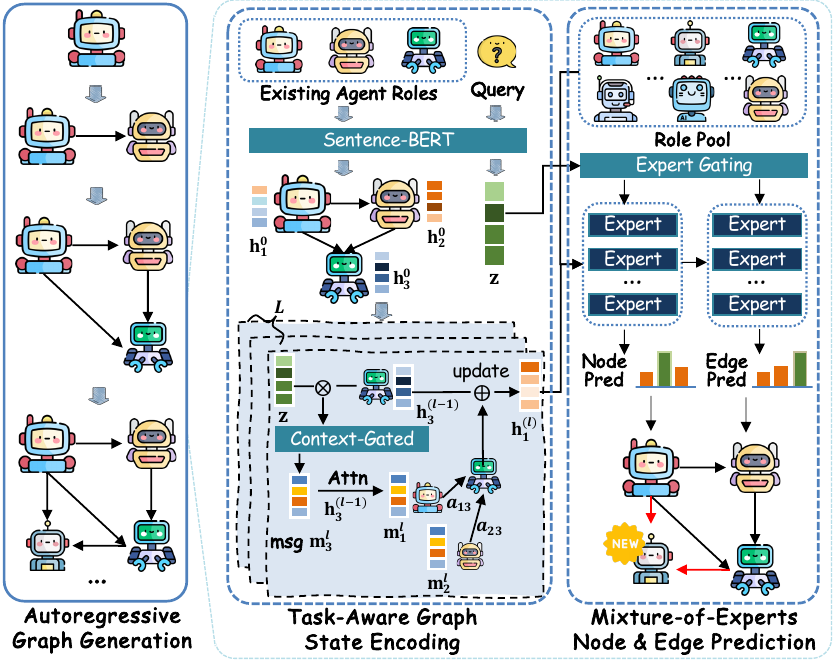}%
  \caption{The workflow of \ourmethod.}
  \label{fig:pipeline}
\end{figure}

In this section, we introduce \ourmethod, a one-for-all MAS topology design approach, with the pipeline demonstrated in Figure~\ref{fig:pipeline}. We employ an \textbf{autoregressive graph generation model} as the backbone of \ourmethod (Section~\ref{sec:backbone}), which balances the performance and flexibility for the MAS design problem. Building upon the backbone, we design two modules to better support the one-for-all topology design: \textbf{task-aware graph state encoder} (Section~\ref{sec:enc}) that maintains task-dependent representations of the partially constructed graph for guiding topology generation, and \textbf{mixture-of-experts generation module} (Section~\ref{sec:moe}) that adaptively routes experts for node and edge generation according to the contextual requirements of different task types. To optimize \ourmethod, we carefully design a \textbf{three-stage training curriculum} with an \textbf{LLM-based data synthesis approach} (Section~\ref{sec:train}). All components are detailed in the following subsections.

\subsection{Autoregressive Graph Generation Backbone}\label{sec:backbone}
Under the ``one-for-all'' paradigm, a single model should serve diverse domains with varying collaboration patterns. Nevertheless, existing graph learning-based models typically fix scale and role allocation, with a limited capability in role adaptation and scale generalization~\cite{zhang2025g}. Specifically, they cannot simultaneously accommodate different structural requirements across domains nor scale to handle task-specific variations within each domain.

To address these limitations, we reframe the topology design problem: instead of adapting fixed configurations, we propose generating customized graphs from scratch. This generative approach naturally aligns with the OFA paradigm, offering greater flexibility for constructing domain-specific structures and handling the scalability demands of diverse tasks. We achieve this by learning a \textbf{conditional graph generation model} $P(\mathcal{G} \mid \mathcal{Q}, \mathcal{R})$ that directly captures the mapping from task queries to effective collaborative topologies. This model aims to find the optimal topology $\mathcal{G}^*$ that maximizes the learned distribution:

\begin{equation}
\mathcal{G}_{\text{gen}}^* = \arg\max_{\mathcal{G} \in \mathbb{G}} P(\mathcal{G} \mid \mathcal{Q}, \mathcal{R}_{\text{OFA}}),
\end{equation}
where $\mathcal{Q}$ is the task query, and $\mathcal{R}_{\text{OFA}}$ denotes the universal role pool spanning all domains.

To make this complex joint distribution $P(\mathcal{G} \mid \mathcal{Q}, \mathcal{R}_{\text{OFA}})$ tractable, we employ an autoregressive graph generation framework that decomposes topology construction into a sequence of local decisions. 
Specifically, the generation process unfolds step-by-step: at each step $t$, we first select an agent role $v_t$, then predict incoming edges $\mathcal{E}_t$ connecting existing agents to this new node. The complete generation probability is factorized as:
\begin{equation}
P(\mathcal{G}|\mathcal{Q}) = \prod_{t=1}^{|\mathcal{V}|} \left[ P(v_t | \mathcal{G}_{<t}, \mathcal{Q}) \cdot P(\mathcal{E}_t | v_t, \mathcal{G}_{<t}, \mathcal{Q}) \right],
\end{equation}
where $\mathcal{G}_{<t}$ represents the partially constructed graph with the first $t-1$ nodes and their interconnections. The edge prediction can be further decomposed as:
\begin{equation}
P(\mathcal{E}_t | v_t, \mathcal{G}_{<t}, \mathcal{Q}) = \prod_{j=1}^{t-1} P(e_{j \rightarrow t} | v_t, \mathcal{G}_{<t}, \mathcal{Q}),
\end{equation}
where $e_{j \rightarrow t}$ is a binary variable indicating whether an edge exists from node $j$ to the newly added node $t$.

At each generation step $t$, our model makes two sequential decisions: \textbf{Role Selection:} Determine the most appropriate agent role $v_t$ to add next, considering the current partial topology $\mathcal{G}_{<t}$ and the task requirements $\mathcal{Q}$. This decision leverages learned patterns about which roles effectively complement existing agent configurations. \textbf{Connection Prediction:} For the newly added agent $v_t$, predict which existing agents should provide it with information by modeling the probability of each potential incoming edge. This captures learned dependencies about information flow patterns that facilitate effective collaboration.

The autoregressive formulation is particularly well-suited for the OFA paradigm, providing the following key advantages: (1)~\textbf{Cross-domain Tractability}: Sequential decomposition of the graph generation process allows the model to organize different roles and connections in a flexible manner, which enables the model to generalize across domains with varying requirements; (2)~\textbf{Variable-Size Graph Modeling}: The autoregressive generation is also flexible in terms of variable generated graph sizes and topologies, which can better fit the tasks with different levels of difficulty and heterogeneous topology demands.

While this autoregressive backbone provides a flexible foundation, realizing effective OFA topology design requires addressing two critical gaps: how to integrate task-specific semantic constraints into graph state encoding (Section~\ref{sec:enc}), and how to enable specialized generation strategies for different task categories without sacrificing universality (Section~\ref{sec:moe}). We address these challenges through two key architectural innovations.

\subsection{Task-Aware Graph State Encoder}\label{sec:enc}
At each step of the autoregressive generation process, the model requires an in-depth and contextualized understanding of the partially constructed graph $\mathcal{G}_{<t}$ to make informed decisions about the next node or edge to add. Traditional graph neural networks (GNNs) typically ignore global task context during their message-passing updates, potentially missing crucial task-specific structural requirements~\cite{pan2025label,chen2025uncertainty,zhao2025freegad}. To address this limitation, we design a specialized encoder that we term the \textbf{Task-Aware Graph State Encoder (TAGSE)}, which persistently integrates the task context into every layer of its message-passing computations. %

\noindent\textbf{Unified Semantic Representation.} 
The foundation of task-aware graph encoding lies in obtaining informative representations of input task queries and role descriptions to guide the graph generation process. Since our OFA model needs to process queries across diverse domains, such as mathematical reasoning and code generation, we require a unified semantic space that enables cross-domain learning while retaining domain-specific features. To this end, we employ a dedicated encoding pipeline where a pre-trained sentence transformer first obtains a fixed-dimensional embedding $\mathbf{e}_{\mathcal{Q}}$ for the given query $\mathcal{Q}$, which is then projected into the final task vector $\mathbf{z} \in \mathbb{R}^{d_{\text{task}}}$ through an MLP-based task encoder. This task vector $\mathbf{z}$ serves as the global conditioning signal that will be integrated into every layer of the subsequent graph encoding process. In addition, we establish node initial states $\mathbf{h}_v^{(0)}$ by encoding agent role descriptions (e.g., \texttt{Analyst} and \texttt{Programmer}) using the same pre-trained sentence transformer, ensuring semantic consistency between task requirements and agent capabilities within the unified representation space. 
Given the initial states and task representation as input, for each layer $l \in \{1, \dots, L\}$, our encoder updates the node hidden states $\mathbf{h}_v^{(l)}$ through the following mechanisms.

\noindent\textbf{Context-Gated Message Passing.} To ensure only task-relevant information flows between nodes, we employ a gating mechanism controlled by the task vector $\mathbf{z}$ that element-wise modulates the message from each node:
\begin{equation}
\mathbf{m}_v^{(l)} = \sigma(\mathbf{W}_g[\mathbf{h}_v^{(l-1)} \| \mathbf{z}]) \odot \text{ReLU}(\mathbf{W}_m \mathbf{h}_v^{(l-1)}),
\end{equation}
where $\mathbf{W}_g$ and $\mathbf{W}_m$ are learnable weight matrices, $[\cdot \| \cdot]$ denotes concatenation, and $\odot$ represents element-wise multiplication. To encourage selective activation and prevent the gate from becoming overly dense, we apply L1 regularization to the gate values $\sigma(\mathbf{W}_g[\mathbf{h}_v^{(l-1)} \| \mathbf{z}])$, promoting sparsity in the gating decisions and ensuring that only the most task-relevant information channels are activated.

\noindent\textbf{Role-Aware Attention Aggregation.} Given that different agent roles may require different types of information, we employ an attention mechanism that allows nodes to prioritize messages based on both the sender's current state and the receiver's processed message:
\begin{equation}
e_{uv}^{(l)} = \text{LeakyReLU}(\mathbf{a}^T [\mathbf{W}_k \mathbf{h}_u^{(l-1)} \| \mathbf{W}_q \mathbf{m}_v^{(l)}]),
\end{equation}
\begin{equation}
\alpha_{uv}^{(l)} = \frac{\exp(e_{uv}^{(l)})}{\sum_{j \in \mathcal{N}_v} \exp(e_{jv}^{(l)})},
\end{equation}
where $\mathbf{W}_k, \mathbf{W}_q,$ and $\mathbf{a}$ are learnable parameters, and $\mathcal{N}_v$ denotes the neighborhood of node $v$.

\noindent\textbf{Residual State Update.} To ensure stable training and preserve important information from previous layers, we employ a residual connection that combines the previous node state with the newly aggregated message information:
\begin{equation}
\mathbf{h}_v^{(l)} = \frac{1}{2} \left( \mathbf{h}_v^{(l-1)} + \hat{\mathbf{m}}_v^{(l)} \right),
\end{equation}
where $\hat{\mathbf{m}}_v^{(l)} = \sum_{u \in \mathcal{N}_v} \alpha_{uv}^{(l)} \mathbf{m}_u^{(l)}$ represents the attention-weighted aggregated message.

The proposed TAGSE encoder deeply integrates task information into the message passing process, leading to sensitivity to queries from different tasks. Meanwhile, during aggregation, it considers both the correlations among connected roles and the ego information of each node state, which generates a comprehensive encoding for each agent. The high-quality embeddings learned by TAGSE provide an essential foundation for generation decisions.

\subsection{Mixture-of-Experts Generation Module}\label{sec:moe}
A single and static GNN often struggles to fit diverse task types, as its parameters have to find a suboptimal compromise between different reasoning patterns. To enable strong versatility in topology generation, we employ a Mixture-of-Experts (MoE) paradigm that allows the model to dynamically reconfigure its computational pathways based on task characteristics. This module consists of two components, namely \textit{expert gating network} and \textit{MoE prediction head}, that work together to provide specialized and task-adaptive generation capabilities.

\subsubsection{Expert Gating Network}
The gating network serves as the decision-making component that determines which specialized experts should be activated for a given task. Taking the task vector $\mathbf{z}$ as input, a lightweight gating network computes a probability distribution over $K$ available expert networks:
\begin{equation}
\mathbf{w} = \text{Softmax}(\text{MLP}_\text{gate}(\mathbf{z})),
\end{equation}
where $\mathbf{w} \in \mathbb{R}^K$ represents the expert activation weights. This mechanism enables the model to learn distinct ``design strategies'' for different categories of tasks. For instance, mathematical reasoning tasks may activate experts that favor sequential chain topologies (e.g., ``Analyst → Solver → Verifier''), while code debugging tasks may prefer hierarchical review structures (e.g., multiple Reviewers feeding into a central Debugger), and the model dynamically combines their insights based on the current task's requirements.

\subsubsection{MoE Prediction Heads}
The following step in the autoregressive generation process leverages the task-aware graph representations and the expert selection weights to make precise predictions of graph structure expansion. The prediction heads operate in two stages corresponding to the two fundamental decisions required at each generation step:

\noindent\textbf{Node Role Prediction.} To determine the next agent role to add to the graph, we employ an MoE layer that processes the global graph representation. Each of the $K$ expert networks produces its own prediction for the role distribution, and the final output combines these predictions using the gating weights:
\begin{equation}
P(\text{role}_t | \mathcal{G}_{<t}, \mathbf{z}) = \sum_{k=1}^{K} w_k \cdot \text{Expert}_k^{\text{node}}(\text{GlobalPool}(\mathcal{G}_{<t}), \mathbf{z}),
\end{equation}
where $\text{GlobalPool}(\mathcal{G}_{<t})$ represents a global representation of the partial graph obtained through pooling operations over node representations.

\noindent\textbf{Edge Connection Prediction.} For each potential incoming edge to the newly added node, another set of expert networks predicts the connection probability. Taking the concatenation of the source node's representation, the target node's representation, and the task vector, each expert generates a prediction for the connectivity, and then the predictions are fused together to form the final probability:
\begin{equation}
P(\text{edge}_{j \rightarrow t} | \mathcal{G}_{<t}, \mathbf{z}) = \sum_{k=1}^{K} w_k \cdot \text{Expert}_k^{\text{edge}}([\mathbf{h}_j \| \mathbf{h}_t \| \mathbf{z}]),
\end{equation}
where $\mathbf{h}_j$ and $\mathbf{h}_t$ represent the node representations of the source and target nodes, respectively.

This MoE-based architecture ensures that every generation decision results from a task-specific ensemble of specialized experts, enabling the model to adapt its generation strategy dynamically while maintaining consistency through the shared task representation.

\begin{table*}[!t]
    \caption{Performance comparison (\%). The best results are highlighted in \textbf{bold}, and the second-best results are \underline{underlined}.}\label{tab:performance}
    \centering
    \small
    \begin{tabular}{l|p{1.7cm}<{\centering}p{1.7cm}<{\centering}p{1.7cm}<{\centering}p{1.7cm}<{\centering}p{1.7cm}<{\centering}p{1.7cm}<{\centering}|p{1.7cm}<{\centering}}
    \toprule
    \textbf{Method} & \textbf{MMLU} & \textbf{GSM8K} & \textbf{AQuA} & \textbf{MultiArith} & \textbf{SVAMP} & \textbf{HumanEval} & \textbf{Average} \\
    \midrule
    \rowcolor{Gray}
        \multicolumn{8}{c}{\textbf{Direct Answer}} \\
    Vanilla                                & 80.39      & 82.30          & 71.06              & 93.09        & 86.55          & 71.39             & 80.80         \\ \midrule
    \rowcolor{Gray}
        \multicolumn{8}{c}{\textbf{Single-Agent Prompting}} \\
    CoT                              & 81.69 \scriptsize$\uparrow$1.30 &
    86.50 \scriptsize$\uparrow$4.20 &
    73.58 \scriptsize$\uparrow$2.52 &
    93.25 \scriptsize$\uparrow$0.16 &
    87.36 \scriptsize$\uparrow$0.81 &
    74.67 \scriptsize$\uparrow$3.28 &
    82.84 \scriptsize$\uparrow$2.04          \\
    SC (CoT)               &
    83.66 \scriptsize$\uparrow$3.27 &
    81.60 \scriptsize$\downarrow$0.70 &
    75.63 \scriptsize$\uparrow$4.57 &
    94.12 \scriptsize$\uparrow$1.03 &
    88.59 \scriptsize$\uparrow$2.04 &
    79.83 \scriptsize$\uparrow$8.44 &
    83.91 \scriptsize$\uparrow$3.11          \\
     \midrule
     \rowcolor{Gray}
        \multicolumn{8}{c}{\textbf{Fixed Multi-Agent Topologies}} \\
     Chain                              &83.01 \scriptsize$\uparrow$2.62 &
    88.30 \scriptsize$\uparrow$6.00 &
    74.05 \scriptsize$\uparrow$2.99 &
    93.27 \scriptsize$\uparrow$0.18 &
    87.17 \scriptsize$\uparrow$0.62 &
    81.37 \scriptsize$\uparrow$9.98 &
    84.53 \scriptsize$\uparrow$3.73          \\
    Tree              &
    81.04 \scriptsize$\uparrow$0.65 &
    85.20 \scriptsize$\uparrow$2.90 &
    71.23 \scriptsize$\uparrow$0.17 &
    93.68 \scriptsize$\uparrow$0.59 &
    88.91 \scriptsize$\uparrow$2.36 &
    80.53 \scriptsize$\uparrow$9.14 &
    83.43 \scriptsize$\uparrow$2.63        \\
    Complete                           &
    82.35 \scriptsize$\uparrow$1.96 &
    80.10 \scriptsize$\downarrow$2.20 &
    72.95 \scriptsize$\uparrow$1.89 &
    94.53 \scriptsize$\uparrow$1.44 &
    84.01 \scriptsize$\downarrow$2.54 &
    79.03 \scriptsize$\uparrow$7.64 &
    82.16 \scriptsize$\uparrow$1.36         \\
    Random               &
    84.31 \scriptsize$\uparrow$3.92 &
    86.90 \scriptsize$\uparrow$4.60 &
    76.48 \scriptsize$\uparrow$5.42 &
    94.08 \scriptsize$\uparrow$0.99 &
    87.54 \scriptsize$\uparrow$0.99 &
    82.66 \scriptsize$\uparrow$11.27 &
    85.33 \scriptsize$\uparrow$4.53        \\
    LLM-Debate                           &
    84.96 \scriptsize$\uparrow$4.57 &
    91.40 \scriptsize$\uparrow$9.10 &
    77.65 \scriptsize$\uparrow$6.59 &
    96.36 \scriptsize$\uparrow$3.27 &
    90.11 \scriptsize$\uparrow$3.56 &
    84.70 \scriptsize$\uparrow$13.31 &
    87.53 \scriptsize$\uparrow$6.73       \\
     \midrule
     \rowcolor{Gray}
        \multicolumn{8}{c}{\textbf{Design Models with Dataset-Specific Training}} \\
    AgentPrune                 & 85.07 \scriptsize$\uparrow$4.57      & 91.10 \scriptsize$\uparrow$8.80          & 80.51 \scriptsize$\uparrow$9.45       & 94.65 \scriptsize$\uparrow$1.56               & 90.58 \scriptsize$\uparrow$4.03          & 86.75 \scriptsize$\uparrow$15.36              & 88.09 \scriptsize$\uparrow$7.29         \\
    AgentDropout              & 85.62 \scriptsize$\uparrow$5.23        & 91.70 \scriptsize$\uparrow$9.40          & 80.94 \scriptsize$\uparrow$9.88        & 95.60 \scriptsize$\uparrow$2.51               & 91.04 \scriptsize$\uparrow$4.49         & 85.98 \scriptsize$\uparrow$14.59              & 88.48 \scriptsize$\uparrow$7.68         \\
    G-designer             & 86.92 \scriptsize$\uparrow$6.53       & 93.80 \scriptsize$\uparrow$11.50         & 81.60 \scriptsize$\uparrow$10.54         & 96.50 \scriptsize$\uparrow$3.41            & 93.10 \scriptsize$\uparrow$6.55          & 88.33 \scriptsize$\uparrow$16.94             & 90.04 \scriptsize$\uparrow$9.24        \\
    EIB-LEARNER             & \underline{88.90} \scriptsize$\uparrow$8.51       & \underline{95.20} \scriptsize$\uparrow$12.90         & \underline{83.49} \scriptsize$\uparrow$12.43         & \underline{96.83} \scriptsize$\uparrow$3.74            & \underline{94.70} \scriptsize$\uparrow$8.15          & 89.15 \scriptsize$\uparrow$17.76             & 91.38 \scriptsize$\uparrow$10.58        \\  \midrule
    \rowcolor{Gray}
        \multicolumn{8}{c}{\textbf{One-for-All Topology Design}} \\
    \ourmethod (pre-trained) & 
    \underline{88.90} \scriptsize$\uparrow$8.51 & 
    94.90 \scriptsize$\uparrow$12.60 & 
    83.18 \scriptsize$\uparrow$12.12 & 
    \textbf{99.11} \scriptsize$\uparrow$6.02 & 
    94.27 \scriptsize$\uparrow$7.72 & 
    \underline{92.56} \scriptsize$\uparrow$21.17 & 
    \underline{92.15} \scriptsize$\uparrow$11.35 \\  
    \ourmethod (fine-tuned) & 
    \textbf{89.54} \scriptsize$\uparrow$9.15 & 
    \textbf{95.30} \scriptsize$\uparrow$13.00 & 
    \textbf{85.05} \scriptsize$\uparrow$13.99 & 
    \textbf{99.11} \scriptsize$\uparrow$6.02 & 
    \textbf{94.90} \scriptsize$\uparrow$8.35 & 
    \textbf{94.21} \scriptsize$\uparrow$22.82 & 
    \textbf{93.02} \scriptsize$\uparrow$14.92 \\ 
    \bottomrule
    \end{tabular}
    \end{table*}

\subsection{Training Strategy and Data Construction}\label{sec:train}
In order to train a versatile model like \ourmethod, a large-scale set of high-quality `(task, optimal\_graph)` supervised data is highly required. Nevertheless, it is costly to collect such annotations due to the intensive human expertise required for structure design and the computational overhead of verifying structural validity. At the same time, directly training on heterogeneous multi-domain data may lead to degraded performance, as the model can suffer from catastrophic forgetting and unstable convergence. 
To overcome these limitations, we devise a three-stage training curriculum that bootstraps the model's knowledge, evolving from general structural understanding to specific and high-quality task alignment. 
During the training process, we introduce an LLM-driven data synthesis module to alleviate the data limitation, enabling us to train \ourmethod on large-scale ``query-topology'' data without human annotation. 
This strategy makes the training of an OFA generative model feasible by systematically building knowledge from low-cost and large-scale data before refining it with a small amount of empirically validated data.

\subsubsection{Stage 1: Unconditional Graph Pre-training}
A randomly initialized generative model has no intrinsic knowledge of what constitutes a valid or efficient graph structure. To address this, we first pre-train the model to learn the fundamental ``grammar'' of communication graphs for MAS, independent of any task. Specifically, we curate a dataset, $\mathcal{D}_{\text{graphs}}$, which includes classic communication topologies (e.g., Chain, Star, and FullConnected) with varying sizes and role allocation, and then train \ourmethod model autoregressively on this data. During this stage, the task vector $\mathbf{z}$ is replaced with a zero vector. This stage teaches the generator to learn how to construct valid and coherent graph structures based solely on the sequence of previous decisions, thereby establishing a strong structural prior before any task-specific knowledge is introduced.

\subsubsection{Stage 2: Task-Topology Alignment via LLM-driven Data Synthesis}
Training an OFA model requires extensive (query, topology) pairs spanning diverse domains to learn the mapping from task requirements to optimal graph structures. However, obtaining such data through actual MAS execution would be prohibitively expensive, as it requires running complete multi-agent systems for each query-topology pair to generate ground-truth labels. To address this challenge, we leverage a LLM as a ``proxy system designer'' to generate massive pseudo-labeled datasets at low cost. Concretely, we prompt the LLM to produce a large amount of diverse ``(task query, MAS configuration)'' pairs belonging to different domains; details are in Appendix~\ref{app:trcfg} 
The synthetic MAS configurations are then converted into a structured graph dataset $\mathcal{D}_{\text{gen}}$. To optimize \ourmethod, we perform conditional pre-training on this synthetic dataset, starting from the model after Stage 1 training. This stage efficiently transfers the general-purpose reasoning ability of the LLM into our specialized model, teaching it the alignment patterns between task semantics and graph structures without requiring expensive MAS executions.

\subsubsection{Stage 3: Supervised Generative Fine-tuning}
While the knowledge acquired from the LLM-generated data is broad, it may not be optimal for the specific scenarios. To obtain high-quality training data, we leverage a small fraction of training data from real-world datasets with predefined classical topology configurations to execute MAS systems and obtain empirically validated performance data. Specifically, for each query in the training sets across all domains, we apply established topological structures (e.g., sequential chains, layered, and star) and execute the corresponding MAS to evaluate their performance. We curate a dataset $\mathcal{D}_{\text{fine-tune}}$ consisting of ``(task query, high-performing topology)'' pairs where the topologies represent proven classical configurations that demonstrate superior performance through actual MAS execution.  Finally, we perform fine-tuning on this empirically grounded dataset.

\subsubsection{Training Objective}
To optimize the autoregressive generation model, we define a loss function that captures the core goals of accurate generation, efficient use of the model's expert capacity, and sparse gating behavior. The model is trained end-to-end in all stages by minimizing the loss:
\begin{equation}
L_{\text{total}} = L_{\text{graph}} + \lambda_{\text{balance}} L_{\text{balance}} + \lambda_{\text{gate}} L_{\text{gate}},
\end{equation}
where $\lambda_{\text{balance}}$ and $\lambda_{\text{gate}}$ are scalar hyperparameters that weight the contribution of each loss term. The primary supervised signal, $L_{\text{graph}}$, is calculated under a teacher forcing manner, and is used in all three training stages. It is composed of a node prediction loss and an edge prediction loss at each step of the generation:
\begin{equation}
L_{\text{graph}} = \sum_{t=1}^{|\mathcal{V}|} (L_{\text{node}}^{(t)} + L_{\text{edge}}^{(t)}),
\end{equation}
where $L_{\text{node}}^{(t)}$ is the standard cross-entropy loss for the new node's role, and $L_{\text{edge}}^{(t)}$ is the binary cross-entropy with logits loss for predicting the incoming edges. The auxiliary loss, $L_{\text{balance}}$, encourages the gating network to use all experts, ensuring the full capacity of the MoE architecture is utilized. The sparse gating regularization term, $L_{\text{gate}}$, applies L1 regularization to the gate activations in TAGSE, promoting selective information flow and preventing overly dense gating patterns that could lead to inefficient computation.

\section{Experiments}
\begin{table}[t]
    \caption{Out-of-distribution generalization on GAIA benchmark (unseen dataset during training).}
    \centering
    \small
    \setlength{\tabcolsep}{1pt}{
    \begin{tabular}{l|p{1.33cm}<{\centering}p{1.33cm}<{\centering}p{1.33cm}<{\centering}|p{1.33cm}<{\centering}}
    \toprule
    \textbf{Method} & \textbf{Level-1} & \textbf{Level-2} & \textbf{Level-3} & \textbf{Average} \\ 
    \midrule
    Complete & 5.66 & 8.14 & 0.00 & 4.60 \\
    Chain & 11.32 & 6.98 & 3.85 & 7.38 \\
    AgentPrune & 9.43 & 8.14 & 0.00 & 5.86 \\
    G-Designer & 5.66 & 6.98 & 3.85 & 5.50 \\
    EIB-LEARNER & 9.43 & 6.98 & 0.00 & 5.47 \\ \midrule
    \ourmethod & \textbf{13.21} & \textbf{9.30} & \textbf{3.85} & \textbf{8.79} \\
    \bottomrule
    \end{tabular}
    }
    \label{tab:unseen}
    \end{table}

\subsection{Experimental Setup}

\noindent\textbf{Datasets.} We evaluate on MMLU~\cite{hendrycks2020measuring}, GSM8K~\cite{cobbe2021training}, AQuA~\cite{ling2017program}, MultiArith~\cite{roy2016solving}, SVAMP~\cite{patel2021nlp}, and HumanEval~\cite{chen2021evaluating}, and test OOD generalization on GAIA~\cite{mialon2023gaia}; details are in Appendix~\ref{app:datasets}.

\noindent\textbf{Baselines.} We compare \ourmethod with individual-agent prompting (CoT~\cite{wei2022chain}, Self-Consistency~\cite{wang2022self}), static topologies, debate-based systems, and one-for-one graph learning-based designers (AgentPrune~\cite{zhang2024agentprune}, AgentDropout~\cite{wang2025agentdropout}, G-Designer~\cite{zhang2025g}, EIB-LEARNER~\cite{shen2025understanding}); details are in Appendix~\ref{app:baselines}.
\noindent\textbf{Evaluation and Implementation.} We report the success rate on each test set, and train a single unified \ourmethod model across all datasets while training one-for-one baselines per domain. Implementation details are in Appendix~\ref{app:implementation}.

\subsection{Main Results}
The main performance comparison is presented in Table \ref{tab:performance}. We conduct the comparison with two versions of \ourmethod: \ding{182}~\ourmethod (pre-trained), which is only trained on synthetic data (Stage 1 and Stage 2) without any real or benchmark-specific supervision; \ding{183}~\ourmethod (fine-tuned), which is trained following the full three-stage curriculum, including fine-tuning on real benchmark data.

According to the results, we have the following observations. \ding{182}~\textbf{\ourmethod achieves a state-of-the-art average score of 93.02\%}, outperforming all baselines by a significant margin. This highlights the effectiveness of our one-for-all approach, which learns to generate specialized collaboration topologies for diverse tasks from a single, unified model.
\ding{183}~\textbf{Fixed topologies and generic prompting show sub-optimal performance.} We can see that fixed MAS topologies such as \textit{Chain} (84.53\%) provide a substantial performance boost, confirming that structured agent collaboration is crucial. However, no single fixed topology excels across all tasks. For instance, \textit{Chain} performs well on sequential tasks like HumanEval, while \textit{Complete} struggles. 
\ding{184}~\textbf{\ourmethod significantly outperforms existing learning-based MAS design methods.} For example, it surpasses the strong EIB-LEARNER baseline by 1.64\% points. 
This result validates that a ``one-for-all'' model, trained across multiple domains, can leverage shared structural knowledge to outperform specialized ``one-for-one'' models. \ding{185}~\textbf{\ourmethod can achieve competitive performance even without fine-tuning on real-world data.} 
\ourmethod (pre-trained) achieves a remarkable average performance of 92.15\%, surpassing the best-performing baseline, EIB-LEARNER (91.38\%). This demonstrates the effectiveness of our synthetic data–based pre-training strategy.

\subsection{Out-of-Distribution Generalization} 
To evaluate generalization to unseen domains, we test all methods on GAIA, which was never seen during training. For fair evaluation, GPT-4o-mini is used as the base model without applying agentic techniques (e.g., tool calling), making the GAIA performance not directly comparable with commercial frameworks. Table~\ref{tab:unseen} presents results across three difficulty levels. We observe: \ding{182}~\textbf{\ourmethod demonstrates strong zero-shot transfer capability}, achieving 8.79\% average accuracy and outperforming all baselines. It excels on Level-1 tasks (13.21\%), showing effective adaptation to new task types. \ding{183}~\textbf{Learning-based baselines struggle with OOD tasks.} G-Designer only achieves 5.50\% average accuracy, underperforming even the simple Chain topology (7.38\%). This validates that our one-for-all paradigm learns more generalizable collaboration patterns than domain-specialized one-for-one approaches.

\begin{figure*}[t]
    \centering
    \begin{subfigure}[b]{0.35\textwidth}
        \centering
        \includegraphics[width=\textwidth]{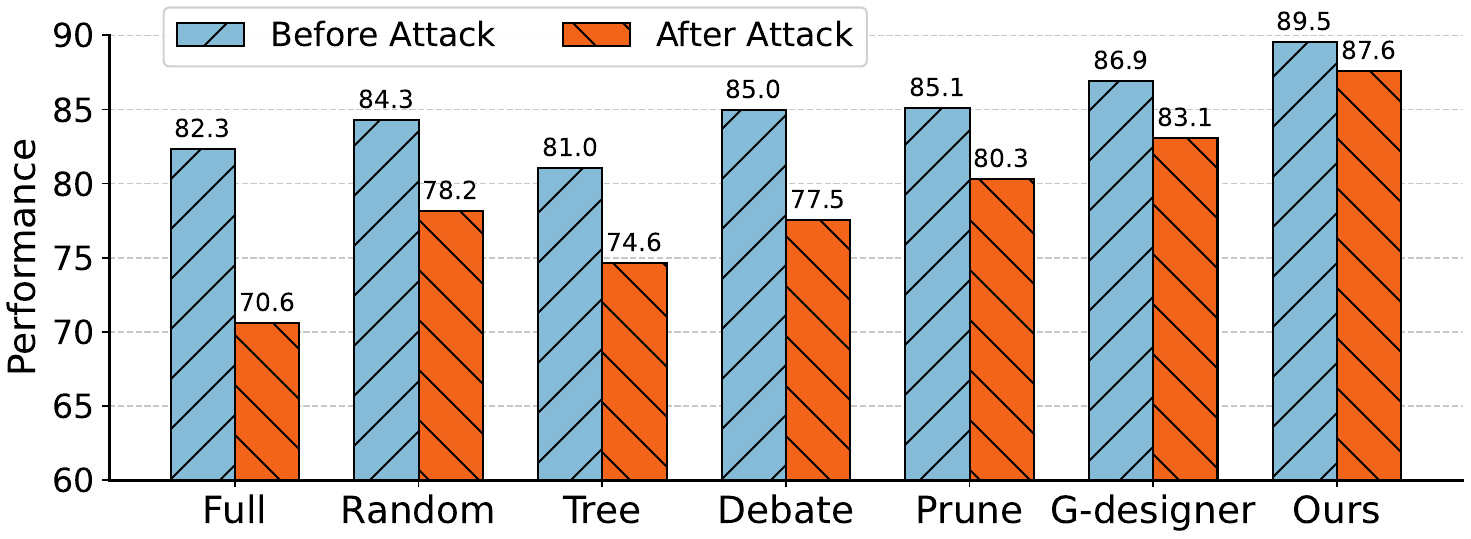}
        \caption{Robustness Analysis}
        \label{fig:robust_sub}
    \end{subfigure}
    \hfill
    \begin{subfigure}[b]{0.2\textwidth}
        \centering
        \includegraphics[width=\textwidth]{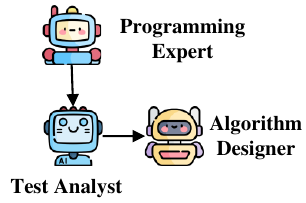}
        \caption{Case: HumanEval}
        \label{fig:case_humaneval}
    \end{subfigure}
    \hfill
    \begin{subfigure}[b]{0.13\textwidth}
        \centering
        \includegraphics[width=\textwidth]{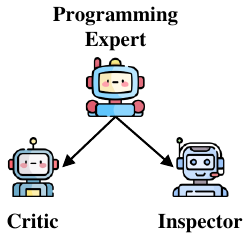}
        \caption{Case: GSM8K}
        \label{fig:case_gsm8k}
    \end{subfigure}
    \hfill
    \begin{subfigure}[b]{0.3\textwidth}
        \centering
        \includegraphics[height=0.43\textwidth]{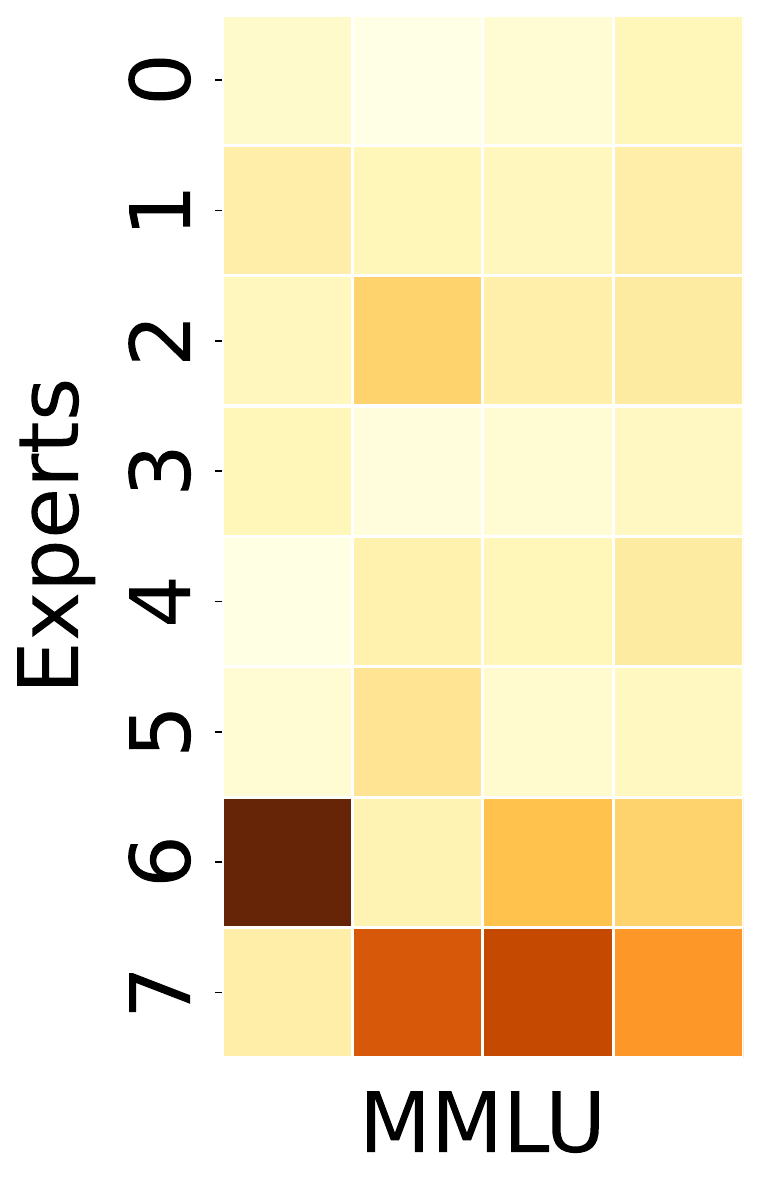}
        \includegraphics[height=0.43\textwidth]{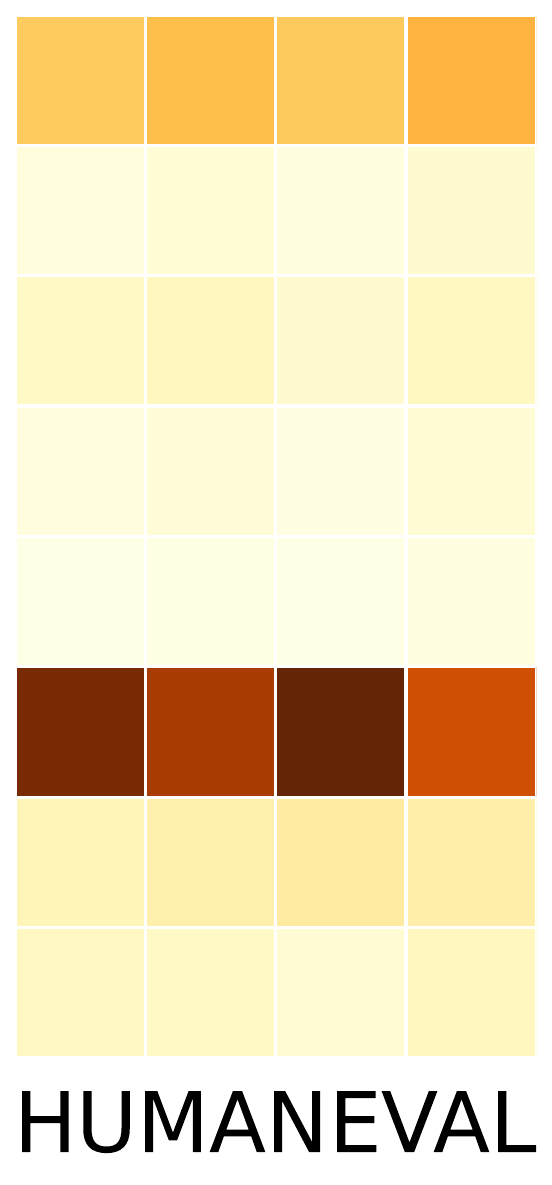}
        \includegraphics[height=0.43\textwidth]{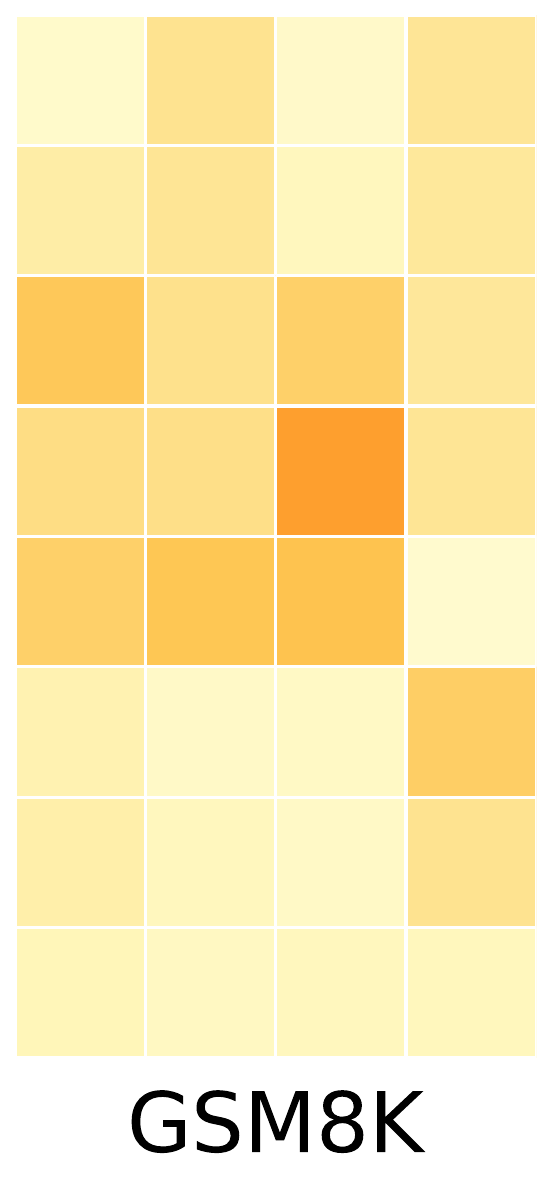}
        \includegraphics[height=0.43\textwidth]{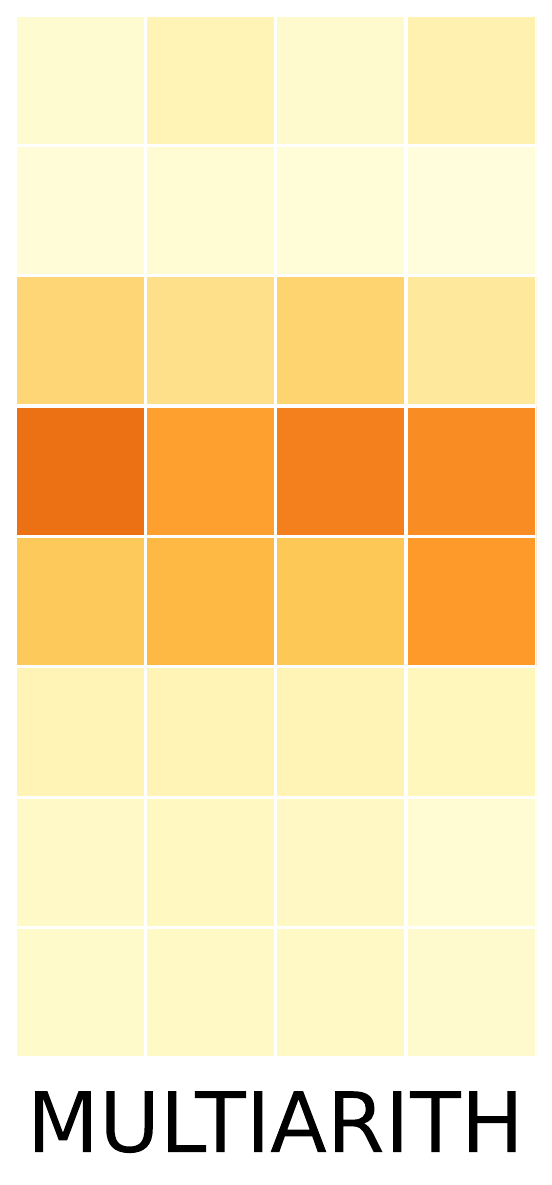}
        \caption{MoE Specialization}
        \label{fig:moe_heatmaps_sub}
    \end{subfigure}
    
    \caption{Comprehensive analysis including (a) robustness against adversarial attacks, (b-c) case study visualizations of generated topologies, and (d) MoE expert activation heatmaps across four domains showing clear specialization patterns.}
    \label{fig:comprehensive_analysis}
\end{figure*}

\begin{table}[t]
    \caption{Results of ablation study.}
    \centering
    \small
    \setlength{\tabcolsep}{1pt}{
    \begin{tabular}{p{1.6cm}|p{1.6cm}<{\centering}p{1.6cm}<{\centering}p{1.6cm}<{\centering}|p{1.6cm}<{\centering}}
    \toprule
    \textbf{Method} & \textbf{MMLU} & \textbf{AQuA} & \textbf{HumanEval} & \textbf{Average} \\ 
    \midrule
    \ourmethod  & \textbf{89.54} & \textbf{85.05} & \textbf{94.21} & \textbf{89.60} \\ \midrule
    w/o MoE & 87.58 & 84.11 & 90.91 & 87.53 \\
    w/o TASE & 88.24 & 83.64 & 90.91 & 87.60 \\
        \midrule
    w/o Stage 1 & 88.24 & 84.58 & 90.08 & 87.08 \\
    w/o Stage 2 & 85.62 & 84.11 & 90.91 & 86.88 \\ \midrule
    w/o $L_{\text{balance}}$ & 88.24 & 83.64 & 88.43 & 86.77 \\
    w/o $L_{\text{gate}}$ & 86.27 & 81.78 & 93.39 & 87.15 \\ 
    \bottomrule
    \end{tabular}
    }
    \label{tab:ablation}
    \end{table}

\subsection{Ablation Study}
To validate the effectiveness of our designs, we conduct a series of ablation studies, with results presented in Table~\ref{tab:ablation}. We have the following observations. \ding{182}~\textbf{Our architectural innovations substantially improve topology generation.} Specifically, replacing these components with default modules leads to a significant performance drop, demonstrating the effectiveness of our tailored designs. \ding{183}~\textbf{Both of the two pre-training stages (i.e., Stage 1 and 2) bring critical performance improvements.} We can see that removing each of them can bring significant degradation, demonstrating their effectiveness in OFA model optimization. \ding{184}~\textbf{The auxiliary losses in \ourmethod also contribute to performance improvement.} As we can see in the results, removing the balance or gate sparsity losses leads to a noticeable drop in accuracy, indicating their role in stabilizing training and enhancing generative learning.

\begin{figure}[t]
  \centering
  \subfloat[Token cost of MMLU]{%
    \label{subfig:tokenMMLU}%
    \includegraphics[width=0.5\columnwidth]{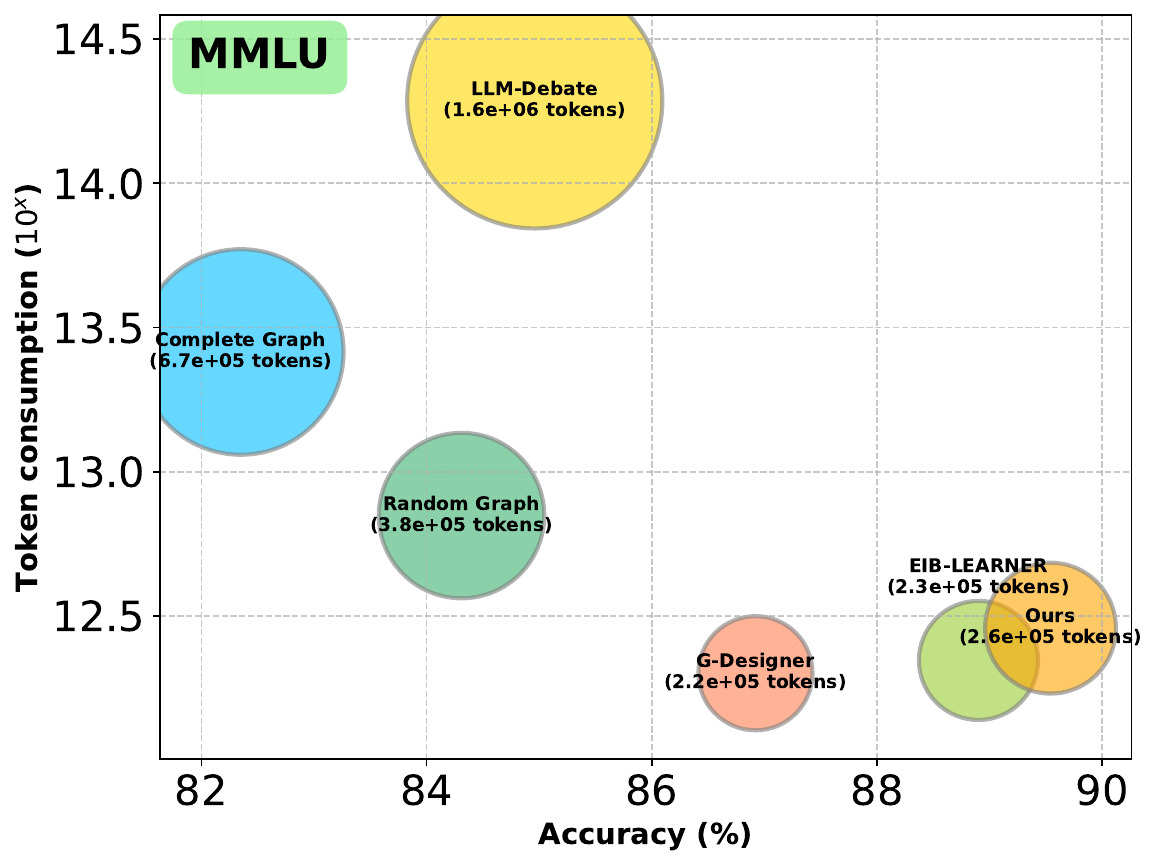}%
  }\hfill
  \subfloat[Token cost of GSM8K]{%
    \label{subfig:tokenGSM8K}%
    \includegraphics[width=0.5\columnwidth]{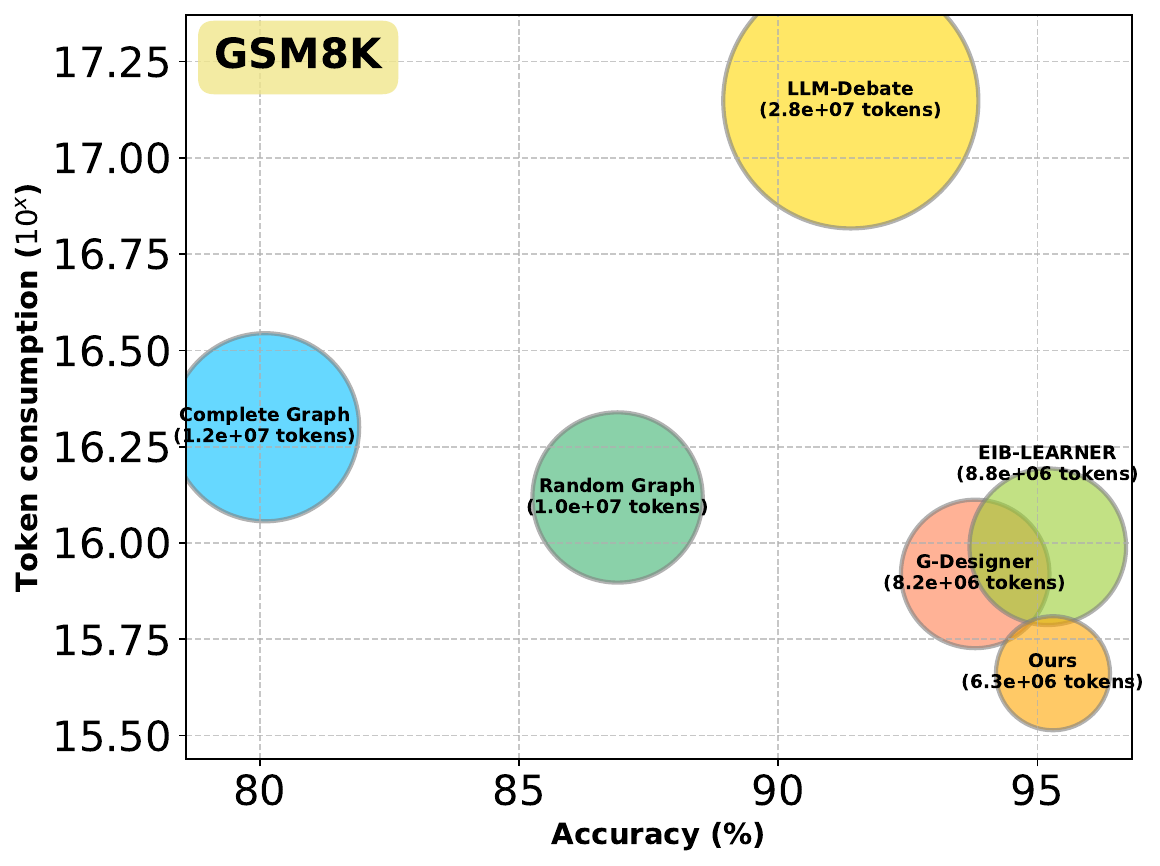}%
  }
  \caption{The prompt token cost comparison.}\label{fig:eff}
\end{figure}

\subsection{Robustness Analysis}
To assess robustness, we evaluate all methods under adversarial conditions by simulating compromised agent scenarios. Specifically, we replace a key agent role (e.g., ``Knowledgeable Exper'') with a malicious prompt designed to generate misleading information that leads to incorrect reasoning.
Figure~\ref{fig:robust_sub} presents the performance comparison before and after adversarial attacks across all baseline methods. We can see that our \textbf{\ourmethod framework demonstrates exceptional resilience}, experiencing only a 2.2\% performance degradation (from 89.54\% to 87.58\%), significantly outperforming all other approaches. In contrast, fixed topology methods show substantial vulnerability, and learning-based methods like G-Designer (4.4\% drop) and AgentPrune (5.6\% drop) demonstrate moderate robustness but remain significantly more vulnerable than our approach.
These results show that \ourmethod maintains this advantage under adversarial scenarios, making it suitable for real-world deployment where agent reliability may be compromised.

\subsection{Analysis of MoE Expert Specialization}
To investigate how the MoE mechanism learns to specialize its experts for different task domains, we visualize the activation weights assigned by the gating network for a sample of tasks, presented in Figure~\ref{fig:moe_heatmaps_sub}. The heatmaps provide compelling evidence by revealing two key phenomena:
\ding{182}~\textbf{Intra-Domain Consistency.} Within a single domain, tasks consistently activate a similar set of experts. For example, across different ``HumanEval'' tasks, the activation patterns are remarkably stable, indicating that the model has learned a consistent and specialized strategy for code generation.
\ding{183}~\textbf{Inter-Domain Specialization.} The preferred set of experts differs sharply across domains, demonstrating clear specialization. For instance, while \textbf{HumanEval} (code generation) heavily relies on a specialized subset (e.g., Experts 0 and 5), \textbf{MMLU} (general knowledge) activates a broader, different combination of experts (e.g., Experts 6 and 7). Similarly, mathematical reasoning domains like \textbf{GSM8K} and \textbf{MultiArith} converge on their own distinct signature, primarily utilizing Experts 2, 3, and 4.

\subsection{Case Study}
Figure~\ref{fig:case_humaneval} and Figure~\ref{fig:case_gsm8k} visualize representative topologies, demonstrating \ourmethod's adaptive role selection. \ding{182}~\textbf{Domain-appropriate role selection.} For the HumanEval coding task (Figure~\ref{fig:case_humaneval}), \ourmethod selects code-generation-specific roles, forming a sequential refinement chain for iterative development. \ding{183}~\textbf{Flexible cross-domain role composition.} For the GSM8K math problem (Figure~\ref{fig:case_gsm8k}), \ourmethod selects math-relevant roles (\texttt{Programming Expert}, \texttt{Inspector}) and incorporates \texttt{Critic}—a role associated with MMLU knowledge tasks—for verification. This shows that \ourmethod selects from a universal role pool rather than domain-specific roles, enabling more flexible collaboration design across task types.

\subsection{Efficiency Analysis}
Figure \ref{fig:eff} illustrates the accuracy-efficiency trade-offs of different methods on MMLU and GSM8K, where bubble size represents token consumption. We observe that \textbf{\ourmethod achieves superior accuracy with competitive efficiency.} On MMLU, \ourmethod attains 89\% accuracy using 2.6e+05 tokens, comparable to learning-based baselines while outperforming them in accuracy. On GSM8K, it achieves 95\% accuracy at the lowest token cost (6.3e+06) among all competing methods. These results demonstrate that \ourmethod not only generates effective collaborative topologies but also maintains computational efficiency.

\section{Conclusion}
In this paper, we introduced \ourmethod, a one-for-all framework for generating adaptive collaboration topologies for multi-agent systems. By combining a task-aware graph state encoder (TAGSE) with a mixture-of-experts (MoE) architecture and a practical three-stage training strategy that progressively builds knowledge from general collaborative graphs to task-specific topologies, our model overcomes the limitations of the traditional one-for-one paradigm. Extensive experiments show that a universal model can not only generalize across diverse domains but can also leverage shared knowledge to achieve state-of-the-art performance. 
This work paves the way for more robust, scalable, and truly autonomous agentic systems, moving from specialized solutions towards a universal graph foundation model for MAS design.

\begin{acks}
The work of S. Pan and Q. Nguyen was partially supported by the Australian Research Council (ARC) under Grant No. DP240101547. The work of Y. Liu was partially supported by the ARC under Grant No. DE260101172. 
\end{acks}

\bibliographystyle{ACM-Reference-Format}
\balance
\bibliography{ref}

@inproceedings{hong2024metagpt,
  title={MetaGPT: Meta programming for a multi-agent collaborative framework},
  author={Hong, Sirui and Zhuge, Mingchen and Chen, Jonathan and Zheng, Xiawu and Cheng, Yuheng and Zhang, Ceyao and Wang, Jinlin and Wang, Zili and Yau, Steven Ka Shing and Lin, Zijuan and others},
  year={2024},
  organization={International Conference on Learning Representations, ICLR}
}

@article{li2026clip,
  title={CLIP-Powered Domain Generalization and Domain Adaptation: A Comprehensive Survey},
  author={Li, Jindong and Li, Yongguang and Fu, Yali and Liu, Jiahong and Liu, Yixin and Yang, Menglin and King, Irwin},
  journal={IEEE Transactions on Pattern Analysis and Machine Intelligence},
  year={2026}
}

@inproceedings{tan2025bisecle,
  title={Bisecle: Binding and Separation in Continual Learning for Video Language Understanding},
  author={Tan, Yue and Hu, Xiaoqian and Xue, Hao and De Melo, Celso and Salim, Flora D},
  booktitle={Advances in Neural Information Processing Systems},
  year={2025}
}

@inproceedings{pan2025survey,
  title={A Survey of Generalization of Graph Anomaly Detection: From Transfer Learning to Foundation Models},
  author={Pan, Junjun and Zheng, Yu and Tan, Yue and Liu, Yixin},
  booktitle={The 16th IEEE International Conference on Knowledge Graphs},
  year={2025}
}

@inproceedings{zhuge2024gptswarm,
  title={Gptswarm: Language agents as optimizable graphs},
  author={Zhuge, Mingchen and Wang, Wenyi and Kirsch, Louis and Faccio, Francesco and Khizbullin, Dmitrii and Schmidhuber, J{\"u}rgen},
  booktitle={International Conference on Machine Learning},
  year={2024}
}

@misc{wang2022self,
      title={Self-Consistency Improves Chain of Thought Reasoning in Language Models}, 
      author={Xuezhi Wang and Jason Wei and Dale Schuurmans and Quoc Le and Ed Chi and Sharan Narang and Aakanksha Chowdhery and Denny Zhou},
      year={2023},
      eprint={2203.11171},
      archivePrefix={arXiv},
      primaryClass={cs.CL},
      url={https://arxiv.org/abs/2203.11171}, 
}

@inproceedings{yao2023react,
  title={React: Synergizing reasoning and acting in language models},
  author={Yao, Shunyu and Zhao, Jeffrey and Yu, Dian and Du, Nan and Shafran, Izhak and Narasimhan, Karthik and Cao, Yuan},
  booktitle={International Conference on Learning Representations (ICLR)},
  year={2023}
}

@article{shinn2023reflexion,
  title={Reflexion: Language agents with verbal reinforcement learning},
  author={Shinn, Noah and Cassano, Federico and Gopinath, Ashwin and Narasimhan, Karthik and Yao, Shunyu},
  journal={Advances in Neural Information Processing Systems},
  volume={36},
  pages={8634--8652},
  year={2023}
}

@inproceedings{zhang2025g,
  title={G-designer: Architecting multi-agent communication topologies via graph neural networks},
  author={Zhang, Guibin and Yue, Yanwei and Sun, Xiangguo and Wan, Guancheng and Yu, Miao and Fang, Junfeng and Wang, Kun and Chen, Tianlong and Cheng, Dawei},
  booktitle={Forty-second International Conference on Machine Learning},
  year={2025}
}

@article{wei2022chain,
  title={Chain-of-thought prompting elicits reasoning in large language models},
  author={Wei, Jason and Wang, Xuezhi and Schuurmans, Dale and Bosma, Maarten and Xia, Fei and Chi, Ed and Le, Quoc V and Zhou, Denny and others},
  journal={Advances in neural information processing systems},
  volume={35},
  pages={24824--24837},
  year={2022}
}

@article{zhang2022automatic,
  title={Automatic chain of thought prompting in large language models},
  author={Zhang, Zhuosheng and Zhang, Aston and Li, Mu and Smola, Alex},
  journal={arXiv preprint arXiv:2210.03493},
  year={2022}
}

@article{yao2023tree,
  title={Tree of thoughts: Deliberate problem solving with large language models},
  author={Yao, Shunyu and Yu, Dian and Zhao, Jeffrey and Shafran, Izhak and Griffiths, Tom and Cao, Yuan and Narasimhan, Karthik},
  journal={Advances in neural information processing systems},
  volume={36},
  pages={11809--11822},
  year={2023}
}

@inproceedings{qian2024scaling-macnet,
  title={Scaling large-language-model-based multi-agent collaboration},
  author={Qian, Chen and Xie, Zihao and Wang, Yifei and Liu, Wei and Dang, Yufan and Du, Zhuoyun and Chen, Weize and Yang, Cheng and Liu, Zhiyuan and Sun, Maosong},
  booktitle={Thirteenth International Conference on Learning Representations},
  year={2025}
}

@inproceedings{wu2024autogen,
  title={AutoGen: Enabling Next-Gen LLM Applications via Multi-Agent Conversation},
  author={Wu, Qingyun and Bansal, Gagan and Zhang, Jieyu and Wu, Yiran and Li, Beibin and Zhu, Erkang and Jiang, Li and Zhang, Xiaoyun and Zhang, Shaokun and Liu, Jiale and others},
  booktitle={ICLR 2024 Workshop on Large Language Model (LLM) Agents},
  year={2024}
}

@article{liu2023dylan,
  title={Dynamic llm-agent network: An llm-agent collaboration framework with agent team optimization},
  author={Liu, Zijun and Zhang, Yanzhe and Li, Peng and Liu, Yang and Yang, Diyi},
  journal={arXiv preprint arXiv:2310.02170},
  year={2023}
}

@inproceedings{du2023improving,
  title={Improving factuality and reasoning in language models through multiagent debate},
  author={Du, Yilun and Li, Shuang and Torralba, Antonio and Tenenbaum, Joshua B and Mordatch, Igor},
  booktitle={International Conference on Machine Learning},
  pages={11733--11763},
  year={2024}
}

@inproceedings{zhang2024agentprune,
  title={Cut the crap: An economical communication pipeline for llm-based multi-agent systems},
  author={Zhang, Guibin and Yue, Yanwei and Li, Zhixun and Yun, Sukwon and Wan, Guancheng and Wang, Kun and Cheng, Dawei and Yu, Jeffrey Xu and Chen, Tianlong},
  booktitle={Thirteenth International Conference on Learning Representations},
  year={2025}
}

@article{wang2025agentdropout,
  title={Agentdropout: Dynamic agent elimination for token-efficient and high-performance llm-based multi-agent collaboration},
  author={Wang, Zhexuan and Wang, Yutong and Liu, Xuebo and Ding, Liang and Zhang, Miao and Liu, Jie and Zhang, Min},
  journal={arXiv preprint arXiv:2503.18891},
  year={2025}
}

@inproceedings{shen2025understanding,
  title={Understanding the Information Propagation Effects of Communication Topologies in LLM-based Multi-Agent Systems},
  author={Shen, Xu and Liu, Yixin and Dai, Yiwei and Wang, Yili and Miao, Rui and Tan, Yue and Pan, Shirui and Wang, Xin},
  booktitle={The Conference on Empirical Methods in Natural Language Processing},
  year={2025}
}

@article{zhu2024survey,
  title={A survey of multi-agent deep reinforcement learning with communication},
  author={Zhu, Changxi and Dastani, Mehdi and Wang, Shihan},
  journal={Autonomous Agents and Multi-Agent Systems},
  volume={38},
  number={1},
  pages={4},
  year={2024},
  publisher={Springer}
}

@article{ning2024survey,
  title={A survey on multi-agent reinforcement learning and its application},
  author={Ning, Zepeng and Xie, Lihua},
  journal={Journal of Automation and Intelligence},
  volume={3},
  number={2},
  pages={73--91},
  year={2024},
  publisher={Elsevier}
}

@inproceedings{besta2024graph,
  title={Graph of thoughts: Solving elaborate problems with large language models},
  author={Besta, Maciej and Blach, Nils and Kubicek, Ales and Gerstenberger, Robert and Podstawski, Michal and Gianinazzi, Lukas and Gajda, Joanna and Lehmann, Tomasz and Niewiadomski, Hubert and Nyczyk, Piotr and others},
  booktitle={Proceedings of the AAAI Conference on Artificial Intelligence},
  volume={38},
  number={16},
  pages={17682--17690},
  year={2024}
}

@inproceedings{khan2024debating,
  title={Debating with more persuasive LLMs leads to more truthful answers},
  author={Khan, Akbir and Hughes, John and Valentine, Dan and Ruis, Laura and Sachan, Kshitij and Radhakrishnan, Ansh and Grefenstette, Edward and Bowman, Samuel R and Rockt{\"a}schel, Tim and Perez, Ethan},
  booktitle={International Conference on Machine Learning},
  pages={23662--23733},
  year={2024}
}

@article{zhang2025evoflow,
  title={EvoFlow: Evolving Diverse Agentic Workflows On The Fly},
  author={Zhang, Guibin and Chen, Kaijie and Wan, Guancheng and Chang, Heng and Cheng, Hong and Wang, Kun and Hu, Shuyue and Bai, Lei},
  journal={arXiv preprint arXiv:2502.07373},
  year={2025}
}

@inproceedings{hu2024self-evomac,
  title={Self-evolving multi-agent collaboration networks for software development},
  author={Hu, Yue and Cai, Yuzhu and Du, Yaxin and Zhu, Xinyu and Liu, Xiangrui and Yu, Zijie and Hou, Yuchen and Tang, Shuo and Chen, Siheng},
  booktitle={Thirteenth International Conference on Learning Representations},
  year={2025}
}

@inproceedings{zhang2025maas,
  title={Multi-agent Architecture Search via Agentic Supernet},
  author={Zhang, Guibin and Niu, Luyang and Fang, Junfeng and Wang, Kun and Bai, Lei and Wang, Xiang},
  booktitle={International Conference on Machine Learning},
  year={2025}
}

@inproceedings{you2018graphrnn,
  title={GraphRNN: Generating Realistic Graphs with Deep Auto-regressive Models},
  author={You, Jiaxuan and Ying, Zhitao and Ren, Xiang and Hamilton, William L and Leskovec, Jure},
  booktitle={International Conference on Machine Learning},
  pages={5694--5703},
  year={2018}
}

@article{liao2019efficient,
  title={Efficient graph generation with graph recurrent attention networks},
  author={Liao, Renjie and Li, Yujia and Song, Yang and Wang, Shenlong and Nash, Charlie and Hamilton, William L and Duvenaud, David and Urtasun, Raquel and Zemel, Richard},
  journal={Advances in Neural Information Processing Systems},
  volume={32},
  year={2019}
}

@inproceedings{mialon2023gaia,
  title={Gaia: a benchmark for general ai assistants},
  author={Mialon, Gr{\'e}goire and Fourrier, Cl{\'e}mentine and Wolf, Thomas and LeCun, Yann and Scialom, Thomas},
  booktitle={The Twelfth International Conference on Learning Representations},
  year={2023}
}

@article{zhao2024pard,
  title={Pard: Permutation-Invariant Autoregressive Diffusion for Graph Generation},
  author={Zhao, Lingxiao and Ding, Xueying and Akoglu, Leman},
  journal={Advances in Neural Information Processing Systems},
  volume={37},
  year={2024}
}

@article{chen2025g2pt,
  title={Graph Generative Pre-trained Transformer},
  author={Chen, Xiaohui and Wang, Yinkai and He, Jiaxing and Du, Yuanqi and Hassoun, Soha and Xu, Xiaolin and Liu, Li-Ping},
  journal={arXiv preprint arXiv:2501.01073},
  year={2025}
}

@article{cohenkarlik2024overcoming,
  title={Overcoming Order in Autoregressive Graph Generation},
  author={Cohen-Karlik, Edo and Rozenberg, Eyal and Freedman, Daniel},
  journal={arXiv preprint arXiv:2402.03387},
  year={2024}
}

@misc{hendrycks2020measuring,
      title={Measuring Massive Multitask Language Understanding}, 
      author={Dan Hendrycks and Collin Burns and Steven Basart and Andy Zou and Mantas Mazeika and Dawn Song and Jacob Steinhardt},
      year={2021},
      eprint={2009.03300},
      archivePrefix={arXiv},
      primaryClass={cs.CY},
      url={https://arxiv.org/abs/2009.03300}, 
}

@misc{cobbe2021training,
      title={Training Verifiers to Solve Math Word Problems}, 
      author={Karl Cobbe and Vineet Kosaraju and Mohammad Bavarian and Mark Chen and Heewoo Jun and Lukasz Kaiser and Matthias Plappert and Jerry Tworek and Jacob Hilton and Reiichiro Nakano and Christopher Hesse and John Schulman},
      year={2021},
      eprint={2110.14168},
      archivePrefix={arXiv},
      primaryClass={cs.LG},
      url={https://arxiv.org/abs/2110.14168}, 
}

@misc{roy2016solving,
      title={Solving General Arithmetic Word Problems}, 
      author={Subhro Roy and Dan Roth},
      year={2016},
      eprint={1608.01413},
      archivePrefix={arXiv},
      primaryClass={cs.CL},
      url={https://arxiv.org/abs/1608.01413}, 
}

@inproceedings{patel2021nlp,
  title={Are NLP Models really able to Solve Simple Math Word Problems?},
  author={Patel, Arkil and Bhattamishra, Satwik and Goyal, Navin},
  booktitle={Proceedings of the 2021 Conference of the North American Chapter of the Association for Computational Linguistics: Human Language Technologies},
  pages={2080--2094},
  year={2021}
}

@misc{ling2017program,
      title={Program Induction by Rationale Generation: Learning to Solve and Explain Algebraic Word Problems}, 
      author={Wang Ling and Dani Yogatama and Chris Dyer and Phil Blunsom},
      year={2017},
      eprint={1705.04146},
      archivePrefix={arXiv},
      primaryClass={cs.AI},
      url={https://arxiv.org/abs/1705.04146}, 
}

@misc{chen2021evaluating,
      title={Evaluating Large Language Models Trained on Code}, 
      author={Mark Chen and Jerry Tworek and Heewoo Jun and Qiming Yuan and Henrique Ponde de Oliveira Pinto and Jared Kaplan and Harri Edwards and Yuri Burda and Nicholas Joseph and Greg Brockman and others},
      year={2021},
      eprint={2107.03374},
      archivePrefix={arXiv},
      primaryClass={cs.LG},
      url={https://arxiv.org/abs/2107.03374}, 
}

@article{liu2025graph,
  title={Graph-augmented large language model agents: Current progress and future prospects},
  author={Liu, Yixin and Zhang, Guibin and Wang, Kun and Li, Shiyuan and Pan, Shirui},
  journal={IEEE Intelligent Systems},
  year={2026}
}

@inproceedings{li2026assemble,
  title={Assemble your crew: Automatic multi-agent communication topology design via autoregressive graph generation},
  author={Li, Shiyuan and Liu, Yixin and Wen, Qingsong and Zhang, Chengqi and Pan, Shirui},
  booktitle={Proceedings of the AAAI Conference on Artificial Intelligence},
  year={2026}
}

@article{liu2024arc,
  title={Arc: A generalist graph anomaly detector with in-context learning},
  author={Liu, Yixin and Li, Shiyuan and Zheng, Yu and Chen, Qingfeng and Zhang, Chengqi and Pan, Shirui},
  journal={Advances in Neural Information Processing Systems},
  volume={37},
  pages={50772--50804},
  year={2024}
}

@inproceedings{li2024noise,
  title={Noise-resilient unsupervised graph representation learning via multi-hop feature quality estimation},
  author={Li, Shiyuan and Liu, Yixin and Chen, Qingfeng and Webb, Geoffrey I and Pan, Shirui},
  booktitle={Proceedings of the 33rd ACM International Conference on Information and Knowledge Management},
  pages={1255--1265},
  year={2024}
}

@article{chen2025uncertainty,
  title={Uncertainty-Aware Graph Neural Networks: A Multihop Evidence Fusion Approach},
  author={Chen, Qingfeng and Li, Shiyuan and Liu, Yixin and Pan, Shirui and Webb, Geoffrey I and Zhang, Shichao},
  journal={IEEE Transactions on Neural Networks and Learning Systems},
  year={2025},
  publisher={IEEE}
}

@inproceedings{zhao2025freegad,
  title={Freegad: A training-free yet effective approach for graph anomaly detection},
  author={Zhao, Yunfeng and Liu, Yixin and Li, Shiyuan and Chen, Qingfeng and Zheng, Yu and Pan, Shirui},
  booktitle={Proceedings of the 34th ACM International Conference on Information and Knowledge Management},
  pages={4379--4389},
  year={2025}
}

@inproceedings{pan2025label,
  title={A label-free heterophily-guided approach for unsupervised graph fraud detection},
  author={Pan, Junjun and Liu, Yixin and Zheng, Xin and Zheng, Yizhen and Liew, Alan Wee-Chung and Li, Fuyi and Pan, Shirui},
  booktitle={Proceedings of the AAAI Conference on Artificial Intelligence},
  volume={39},
  number={12},
  pages={12443--12451},
  year={2025}
}

@article{miao2025blindguard,
  title={Blindguard: Safeguarding llm-based multi-agent systems under unknown attacks},
  author={Miao, Rui and Liu, Yixin and Wang, Yili and Shen, Xu and Tan, Yue and Dai, Yiwei and Pan, Shirui and Wang, Xin},
  journal={arXiv preprint arXiv:2508.08127},
  year={2025}
}

@article{pan2025explainable,
  title={Explainable and Fine-Grained Safeguarding of LLM Multi-Agent Systems via Bi-Level Graph Anomaly Detection},
  author={Pan, Junjun and Liu, Yixin and Miao, Rui and Ding, Kaize and Zheng, Yu and Nguyen, Quoc Viet Hung and Liew, Alan Wee-Chung and Pan, Shirui},
  journal={arXiv preprint arXiv:2512.18733},
  year={2025}
}

@inproceedings{pan2026correcting,
  title={Correcting False Alarms from Unseen: Adapting Graph Anomaly Detectors at Test Time},
  author={Pan, Junjun and Liu, Yixin and Zhou, Chuan and Xiong, Fei and Liew, Alan Wee-Chung and Pan, Shirui},
  booktitle={Proceedings of the AAAI Conference on Artificial Intelligence},
  year={2026}
}


\section*{Appendix}
\appendix

\section{Related Work}\label{app:related_work}

\subsection{Multi-Agent Systems Topology Design}
Multi-agent systems (MAS) have evolved from traditional reinforcement learning-based coordination~\cite{zhu2024survey,ning2024survey,tan2025bisecle} to sophisticated LLM-based collaboration frameworks, where topology design becomes crucial for effective task execution. Early approaches relied on manually crafted, domain-specific templates: sequential chains~\cite{wei2022chain,zhang2022automatic} for step-by-step reasoning, tree-of-thought hierarchies~\cite{yao2023tree,besta2024graph} for systematic exploration, and debate structures~\cite{du2023improving,liu2023dylan,khan2024debating} for adversarial verification. While effective within their respective domains, these fixed patterns lack adaptability across diverse task requirements.

Recent advances have introduced learnable topology generation. Graph-based approaches such as AgentPrune~\cite{zhang2024agentprune} and AgentDrop~\cite{wang2025agentdropout} construct topologies by pruning dense communication graphs, while GPTSwarm~\cite{zhuge2024gptswarm} and G-Designer~\cite{zhang2025g} employ graph neural networks for direct structure generation. Evolutionary methods~\cite{zhang2025evoflow,hu2024self-evomac} and neural architecture search~\cite{zhang2025maas} have also shown promise in discovering optimal agent configurations. However, these approaches universally adopt the ``one-for-one'' paradigm, training specialized models for individual domains, which limits scalability and cross-domain knowledge transfer. Our work addresses this by proposing a unified ``one-for-all'' model with MoE-based specialization.

\subsection{Autoregressive Graph Generation}
Graph generation has evolved through multiple paradigms, with autoregressive methods emerging as a dominant approach due to their natural compatibility with sequential modeling. Early work like GraphRNN~\cite{you2018graphrnn} established the foundation by generating graphs through sequential node and edge addition, while subsequent methods improved efficiency and expressiveness through better graph representations and neural architectures~\cite{liao2019efficient}.

Recent advances have significantly expanded the capabilities of autoregressive graph generation. PARD~\cite{zhao2024pard} introduces permutation-invariant autoregressive diffusion that maintains effectiveness while addressing ordering sensitivity. Graph Generative Pre-trained Transformer (G2PT)~\cite{chen2025g2pt} leverages transformer architectures for autoregressive graph generation with strong generalization capabilities. Methods like Orderless Regularization~\cite{cohenkarlik2024overcoming} tackle the fundamental ordering problem in sequential graph generation by encouraging invariance to different valid orderings.

\section{Dataset Details}\label{app:datasets}

\subsection{In-Distribution Benchmarks}
We evaluate \ourmethod on six diverse benchmarks spanning multiple reasoning domains:

\noindent\textbf{MMLU~\cite{hendrycks2020measuring}.} The Massive Multitask Language Understanding benchmark consists of 57 subjects across STEM, humanities, social sciences, and more. We use a subset of 153 test questions. Each question is multiple-choice with 4 options.

\noindent\textbf{GSM8K~\cite{cobbe2021training}.} A dataset of 8.5K high-quality linguistically diverse grade school math word problems. We use 1,000 problems from the test split. Each problem requires multi-step arithmetic reasoning.

\noindent\textbf{AQuA~\cite{ling2017program}.} Algebra Question Answering dataset contains 254 algebraic word problems with natural language rationales. Problems require algebraic reasoning with multiple solution steps. Each question is multiple-choice with 5 options.

\noindent\textbf{MultiArith~\cite{roy2016solving}.} A collection of 600 elementary school-level arithmetic word problems requiring multi-step reasoning with numerical answers.

\noindent\textbf{SVAMP~\cite{patel2021nlp}.} Simple Variations on Arithmetic Math word Problems contains 1,000 problems with simple variations to test robustness. Problems are designed with subtle linguistic variations to challenge model generalization.

\noindent\textbf{HumanEval~\cite{chen2021evaluating}.} A code generation benchmark with 164 programming problems. Each problem includes a function signature, docstring, body, and unit tests. Evaluated using Pass@1 metric.

\subsection{Out-of-Distribution Benchmark}
\noindent\textbf{GAIA.} The General AI Assistants benchmark is designed to evaluate AI assistants on real-world questions requiring multi-modal reasoning, tool use, and web search. We evaluate on three difficulty levels: Level-1 (53 questions), Level-2 (86 questions), and Level-3 (26 questions). These levels differ in complexity and reasoning requirements. Importantly, GAIA was never seen during training, making it a pure OOD generalization test.

\noindent\textbf{Evaluation Protocol for GAIA.} We adopt different evaluation protocols for one-for-all and one-for-one methods to ensure fair comparison:
\begin{itemize}[leftmargin=*]
    \item \textbf{One-for-All (\ourmethod):} We directly apply our unified model (trained on six in-distribution benchmarks) to GAIA without any additional training, demonstrating true zero-shot OOD generalization capability.
    \item \textbf{One-for-One Baselines:} Given GAIA's knowledge-intensive nature and multi-modal reasoning requirements, we select the baseline models trained on MMLU (which shares similar knowledge-based characteristics) for GAIA evaluation. This provides baselines with the most relevant role configurations and domain knowledge for the OOD task.
\end{itemize}
This protocol highlights the key advantage of our one-for-all paradigm: a single unified model can generalize to unseen domains, while one-for-one methods require careful selection of which domain-specific model to apply.

\subsection{Data Statistics}
Table~\ref{tab:data_stats} provides comprehensive statistics of all datasets used in our experiments, including answer types, evaluation metrics, and licensing information.

\begin{table*}[ht]
    \centering
    \small
    \setlength{\tabcolsep}{3pt}{
     \caption{Dataset statistics and details.} \label{tab:data_stats}

    \begin{tabular}{l|l|ccccl}
    \toprule
    \textbf{Category} & \textbf{Dataset} & \textbf{Answer Type} & \textbf{Metric} & \textbf{\#Train} & \textbf{\#Test} & \textbf{License} \\
    \midrule
    General reasoning & MMLU & Multi-choice & Acc. & 40 & 153 & MIT License \\
    \midrule
    \multirow{4}{*}{Math reasoning} 
    & GSM8K & Number & Acc. & 40 & 1,000 & MIT License \\
    & MultiArith & Number & Acc. & 40 & 560 & Unspecified \\
    & SVAMP & Number & Acc. & 40 & 960 & MIT License \\
    & AQuA & Multi-choice & Acc. & 40 & 214 & Apache-2.0 \\
    \midrule
    Code generation & HumanEval & Code & Pass@1 & 40 & 121 & MIT License \\
    \midrule
    \multirow{3}{*}{OOD} 
    & GAIA (L1) & Multi-modal & Acc. & - & 53 & CC BY 4.0 \\
    & GAIA (L2) & Multi-modal & Acc. & - & 86 & CC BY 4.0 \\
    & GAIA (L3) & Multi-modal & Acc. & - & 26 & CC BY 4.0 \\
    \bottomrule
    \end{tabular}
    }
\end{table*}

\section{Baseline Details}\label{app:baselines}

\noindent\textbf{Single-Agent Prompting Methods.} We compare against three single-agent baselines: (1) \textbf{Vanilla}: Direct prompting without reasoning enhancement; (2) \textbf{Chain-of-Thought (CoT)~\cite{wei2022chain}}: Prompts the LLM to generate intermediate reasoning steps with "Let's think step by step"; (3) \textbf{Self-Consistency (SC)~\cite{wang2022self}}: Samples 5 reasoning paths using CoT and selects the most consistent answer through majority voting.

\noindent\textbf{Fixed Multi-Agent Topologies.} We evaluate five hand-crafted topologies: (1) \textbf{Chain}: Sequential linear chain $A_1 \rightarrow A_2 \rightarrow \cdots \rightarrow A_n$; (2) \textbf{Tree}: Hierarchical structure with multiple layers culminating in a root agent; (3) \textbf{Complete}: Fully connected graph enabling maximum information sharing; (4) \textbf{Random}: Randomly generated topology with random role assignments; (5) \textbf{LLM-Debate}: Debate-based topology where agents critique each other's reasoning over multiple rounds.

\noindent\textbf{Learning-Based Topology Design (One-for-One).} We compare against four SOTA learning-based methods, each trained separately per dataset: 
(1) \textbf{AgentPrune~\cite{zhang2024agentprune}}: Identifies and addresses communication redundancy by performing one-shot pruning on spatial-temporal message-passing graphs, yielding token-economic topologies while defending against adversarial attacks; 
(2) \textbf{AgentDropout~\cite{wang2025agentdropout}}: Inspired by dynamic role adjustment in management theory, identifies redundant agents and communications across different rounds by optimizing adjacency matrices, eliminating them to enhance token efficiency and task performance; 
(3) \textbf{G-Designer~\cite{zhang2025g}}: Employs a variational graph auto-encoder to encode agents and task-specific virtual nodes, dynamically designing task-adaptive communication topologies customized for each task's difficulty and requirements; 
(4) \textbf{EIB-LEARNER~\cite{shen2025understanding}}: Uses a causal framework to analyze error propagation patterns and designs topologies by fusing connectivity patterns from both dense and sparse graphs, balancing error suppression with beneficial information diffusion.

\section{Implementation Details}\label{app:implementation}

\subsection{Model Architecture}
\noindent\textbf{Task Encoder.} We use Sentence-BERT (all-MiniLM-L6-v2) as the pre-trained sentence transformer, which outputs 384-dimensional embeddings. The task encoder MLP projects these to $d_{\text{task}} = 128$ dimensions through a 2-layer network with hidden dimension 256.

\noindent\textbf{TAGSE.} The Task-Aware Graph State Encoder uses $L = 4$ layers with hidden dimension $d_h = 256$. Each layer employs context-gated message passing with role-aware attention aggregation. Node features are initially projected from 384 dimensions to 256 dimensions.

\noindent\textbf{MoE Module.} We use $K = 8$ expert networks. Each expert consists of a 2-layer MLP with hidden dimension 256 and ReLU activation. The gating network is a 2-layer MLP: 128 $\rightarrow$ 256 $\rightarrow$ 8 with ReLU and Softmax.

\noindent\textbf{Role Pool.} Our universal role pool $\mathcal{R}_{\text{OFA}}$ contains 19 diverse roles spanning multiple domains, including: Mathematician, Psychologist, Doctor, Algorithm Designer, Programming Expert, Inspector, etc. Additionally, we define START and END tokens for autoregressive generation. This is significantly larger than domain-specific role pools used in one-for-one methods (typically 4-6 roles per domain).

\subsection{Training Configuration}\label{app:trcfg}

\noindent\textbf{Stage 1: Unconditional Pre-training.}
\begin{itemize}[leftmargin=*]
    \item Dataset: $\mathcal{D}_{\text{graphs}}$ with 800 canonical graph topologies
    \item Topologies: Chain, Star, Tree, Layered, Complete (various sizes: 3-10 nodes)
    \item Epochs: 50
    \item Batch size: 32
    \item Learning rate: $2 \times 10^{-3}$ (0.002)
    \item Optimizer: Adam
    \item Scheduler: ReduceLROnPlateau (factor=0.5, patience=10)
\end{itemize}

\noindent\textbf{Stage 2: LLM-Guided Conditional Training.}
\begin{itemize}[leftmargin=*]
    \item Dataset: $\mathcal{D}_{\text{gen}}$ with ~1,000 synthetic (query, topology) pairs
    \item LLM for data synthesis: GPT-4o-mini with temperature 0.7
    \item Domains: Knowledge QA, Math reasoning, Code generation, Logical reasoning, etc.
    \item Initialization: Load pre-trained weights from Stage 1
    \item Epochs: 50
    \item Batch size: 32
    \item Learning rate: $2 \times 10^{-3}$ (0.002)
    \item Optimizer: Adam
    \item Scheduler: ReduceLROnPlateau (factor=0.5, patience=10)
\end{itemize}

\noindent\textbf{Stage 3: Supervised Fine-tuning.}
\begin{itemize}[leftmargin=*]
    \item Dataset: $\mathcal{D}_{\text{fine-tune}}$ with 1,000 empirically validated (query, topology) pairs (40 queries per benchmark $\times$ 6 benchmarks)
    \item Initialization: Load pre-trained weights from Stage 2
    \item Classical topologies tested: Chain, Layered, Star, Mesh, FullConnected
    \item Selection criterion: Top-2 performing topologies per query based on validation accuracy
    \item Epochs: 50
    \item Batch size: 32
    \item Learning rate: $5 \times 10^{-4}$ (0.0005)
    \item Optimizer: Adam
    \item Scheduler: ReduceLROnPlateau (factor=0.5, patience=10)
\end{itemize}

\begin{tcolorbox}[colback=gray!10, colframe=gray!50, title=Core Generation Instructions (Abbreviated)]
\small
\textbf{Query Diversity:} Generate queries across domains including code implementation, mathematical reasoning, multiple-choice questions, knowledge-based Q\&A, medical consultation, and psychology.

\textbf{Balanced Configuration:}
\begin{itemize}[leftmargin=*]
    \item Agent count: Evenly distribute configurations for 2-6 agents
    \item Topology variety: Use all five topology types (Chain, Star, Mesh, FullConnected, Layered)
    \item Comprehensive role coverage: Use all 19 roles from the universal role pool
\end{itemize}

\textbf{DAG Constraint:} All generated topologies must be Directed Acyclic Graphs (no cycles allowed).

\textbf{Redundancy Avoidance:}
\begin{itemize}[leftmargin=*]
    \item For 2 agents: Only use Chain topology
    \item For 3 agents: Do NOT use FullConnected (identical to Mesh)
    \item For $\geq$4 agents: All five topologies are distinct
\end{itemize}

\textbf{Output Format:} JSONL (one JSON object per line) with fields: query, agent\_count, roles, topology, edges.
\end{tcolorbox}

\subsection{Loss Function Hyperparameters}
The total training loss is a weighted combination of three components:
\begin{itemize}[leftmargin=*]
    \item MoE balance loss weight: $\lambda_{\text{balance}} = 0.2$
    \item Gate sparsity (L1) loss weight: $\lambda_{\text{gate}} = 0.1$
\end{itemize}
Total loss: $\mathcal{L} = \mathcal{L}_{\text{graph}} + \lambda_{\text{balance}} \mathcal{L}_{\text{balance}} + \lambda_{\text{gate}} \mathcal{L}_{\text{gate}}$

\subsection{Inference Configuration}
\begin{itemize}[leftmargin=*]
    \item Maximum graph size: 6 nodes (acts as safeguard to prevent over-generation)
    \item Minimum graph size: 2 nodes (END token sampling suppressed until this threshold)
    \item Generation strategy: Greedy decoding (selecting highest probability at each step) or multinomial sampling
    \item Early stopping: Generation terminates when model predicts END token
    \item Edge generation: Probabilistic sampling with Bernoulli distribution; if no edges sampled, force connection to highest-probability node
    \item MAS execution: $K = 1$ rounds of inter-agent communication
    \item Base LLM for agents: GPT-4o with temperature 0.0
\end{itemize}

\subsection{Computational Resources}
The experiments are conducted on a machine with 12 vCPU Intel(R) Xeon(R) Platinum 8255C CPU @ 2.50GHz and a single RTX 2080 Ti with 11GB GPU memory. The operating system of the machine is Ubuntu 20.04. As for software versions, we use Python 3.8, Pytorch 2.1.2, and CUDA 12.1.

\section{LLM Prompt Templates}\label{app:prompts}

\subsection{Prompt for LLM-Driven Data Synthesis (Stage 2)}
We use GPT-4o with temperature 0.7 to generate diverse (query, topology) pairs. The system prompt instructs the LLM to generate queries spanning multiple domains (code implementation, mathematical reasoning, multiple-choice questions, knowledge-based Q\&A, medical consultation, psychology) and propose suitable MAS configurations using roles from our universal role pool and topologies from \{Chain, Star, Mesh, FullConnected, Layered\}. Key instructions include:

\noindent\textbf{Example Output:}
\begin{verbatim}
{"query": "def odd_count(lst)...", "agent_count": 4,
 "roles": ["Project Manager", "Algorithm Designer",
           "Programming Expert", "Test Analyst"],
 "topology": "Chain", "edges": "0->1 1->2 2->3"}
\end{verbatim}

\subsection{Agent Role Prompts}
Our universal role pool contains 19 specialized roles. Representative role prompts include:

\begin{tcolorbox}[colback=blue!5, colframe=blue!50, title=Knowlegable Expert]
\small
You are a knowlegable expert in question answering. Please give several key entities that need to be searched in wikipedia to solve the problem. If there is no entity in the question that needs to be searched in Wikipedia, you don't have to provide it.
\end{tcolorbox}

\begin{tcolorbox}[colback=blue!5, colframe=blue!50, title=Mathematician]
\small
You are a mathematician who is good at math games, arithmetic calculation, and long-term planning.
\end{tcolorbox}

\begin{tcolorbox}[colback=blue!5, colframe=blue!50, title=Algorithm Designer]
\small
You are an algorithm designer. You need to specify the specific design of the algorithm, including the classes that may be defined and the functions used. When the implementation logic is complex, you can give the pseudocode logic of the main algorithm. I hope your reply will be more concise. Preferably within fifty words.
\end{tcolorbox}

\noindent The complete role pool includes: Knowledgeable Expert, Critic, Mathematician, Psychologist, Historian, Doctor, Project Manager, Algorithm Designer, Programming Expert, Test Analyst, Bug Fixer, Math Solver, Mathematical Analyst, Programming Expert for Math, Inspector, and domain-specific variants for choice questions.
\end{document}